\documentclass[aps,prd,twocolumn,amssymb,eqsecnum,showpacs,showkeyes,secnumarabic,graphics,floatfix,nofootinbib,tightenlines,longbibliography]{revtex4-1}
\usepackage{graphicx}
\usepackage{bm}
\usepackage{cancel}
\usepackage{epsfig}
\usepackage{dcolumn}
\usepackage{amsmath}
\usepackage{hhline}
\usepackage{enumerate}
\usepackage[utf8]{inputenc}
\usepackage[dvipsnames]{xcolor}
\usepackage[breaklinks=true,colorlinks=true,
linkcolor=blue,urlcolor=Blue,citecolor=MidnightBlue,% PDF VIEW
bookmarks=true,bookmarksopenlevel=2]{hyperref}
\usepackage{bm}
\usepackage{amsmath}
\usepackage{longtable}

\begin{document}
\newcommand{\newc}{\newcommand}

\newc{\be}{\begin{equation}}
\newc{\ee}{\end{equation}}
\newc{\ba}{\begin{eqnarray}}
\newc{\ea}{\end{eqnarray}}
\newc{\bea}{\begin{eqnarray*}}
\newc{\eea}{\end{eqnarray*}}
\newc{\D}{\partial}
\newc{\ie}{{\it i.e.} }
\newc{\eg}{{\it e.g.} }
\newc{\etc}{{\it etc.} }
\newc{\etal}{{\it et al.}}
\newc{\lcdm}{$\Lambda$CDM }
\newc{\lcdmnospace}{$\Lambda$CDM}
\newc{\wcdm}{$w$CDM }
\newc{\plcdm}{Planck18/$\Lambda$CDM }
\newc{\plcdmnospace}{Planck18/$\Lambda$CDM}
\newc{\omom}{$\Omega_{0m}$ }
\newc{\omomnospace}{$\Omega_{0m}$}
\newcommand{\nn}{\nonumber}
\newc{\ra}{\Rightarrow}
\newc{\baodv}{$\frac{D_V}{r_s}$ }
\newc{\baodvnospace}{$\frac{D_V}{r_s}$}
\newc{\baoda}{$\frac{D_A}{r_s}$ } 
\newc{\baodanospace}{$\frac{D_A}{r_s}$}
\newc{\baodh}{$\frac{D_H}{r_s}$ }
\newc{\baodhnospace}{$\frac{D_H}{r_s}$}

\title{A $w-M$ phantom transition at $z_t<0.1$ as a resolution of the Hubble tension}

\author{George Alestas}\email{g.alestas@uoi.gr}
\affiliation{Department of Physics, University of Ioannina, GR-45110, Ioannina, Greece}
\author{Lavrentios Kazantzidis}\email{l.kazantzidis@uoi.gr}
\affiliation{Department of Physics, University of Ioannina, GR-45110, Ioannina, Greece}
\author{Leandros Perivolaropoulos}\email{leandros@uoi.gr}
\affiliation{Department of Physics, University of Ioannina, GR-45110, Ioannina, Greece}

\date{\today}

\begin{abstract}
A rapid phantom transition of the dark energy equation of state parameter $w$ at a transition redshift $z_t<0.1$ of the form $w(z)=-1+\Delta w\;\Theta (z_t-z)$ with $\Delta w<0$ can lead to a higher value of the Hubble constant while closely mimicking a Planck18/$\Lambda$CDM form of the comoving distance  $r(z)=\int_0^z\frac{dz'}{H(z')}$ for $z>z_t$. Such a transition however would imply a significantly lower value of the SnIa absolute magnitude $M$ than the value $M_C$ imposed by local Cepheid calibrators at $z<0.01$. Thus, in order to resolve the $H_0$ tension it would need to be accompanied by a similar transition in the value of the SnIa absolute magnitude $M$ as $M(z)=M_C+\Delta M \;\Theta (z-z_t)$ with $\Delta M<0$. This is a Late $w-M$ phantom transition ($LwMPT$). It may be achieved by a sudden reduction of the value of the normalized effective Newton constant $\mu=G_{\rm{eff}}/G_{\rm{N}}$ by about $6\%$ assuming that the absolute luminosity of SnIa is proportional to the Chandrasekhar mass which varies as $\mu^{-3/2}$. We demonstrate that such an ultra low $z$ abrupt feature of $w-M$ provides a better fit to cosmological data compared to smooth late time deformations of $H(z)$ that also address the Hubble tension. For $z_t=0.02$ we find $\Delta w\simeq -4$, $\Delta M \simeq -0.1$. This model also addresses the growth tension due to the predicted lower value of $\mu$ at $z>z_t$. A prior of $\Delta w=0$ (no $w$ transition) can still resolve the $H_0$ tension with a larger amplitude $M$ transition with $\Delta M\simeq -0.2$ at $z_t\simeq 0.01$. This implies a larger reduction of $\mu$ for $z>0.01$ (about $12\%$). The $LwMPT$ can be generically induced by a scalar field non-minimally coupled to gravity with no need of a screening mechanism since in this model $\mu=1$ at $z<0.01$.
\end{abstract}
\maketitle

\section{Introduction}
\label{sec:Introduction}
The cosmological comoving distance to redshift $z$ defined in a flat universe as $r(z)=\int_0^z\frac{dz'}{H(z')}$ where $H(z)$  is the Hubble expansion rate has been constrained to a level of about $2\%$ using standard candles (SnIa calibrated with Cepheid stars \cite{Riess:2019cxk,Riess:2018byc,Riess:2016jrr} and Red Giant stars \cite{Freedman:2020dne,Freedman:2019jwv} or megamasers in accretion disks \cite{Pesce:2020xfe}), a standard ruler (the sound horizon at last scattering calibrated using the CMB anisotropy spectrum \cite{Aghanim:2018eyx,Addison:2017fdm} and/or Big Bang Nucleosynthesis (BBN) \cite{Schoneberg:2019wmt}), strong gravitational lensing \cite{Birrer:2018vtm,Shajib:2019toy} and gravitational waves \cite{Abbott:2019yzh,Soares-Santos:2019irc}. The comoving distance determined using calibrated standard candles  at $z<0.1$  is offset with the comoving distance determined using the sound horizon standard ruler at $z>0.1$, by about $9\%$ which corresponds to a tension of about $5\sigma$. In the context of a \plcdm form of $H(z)$ this mismatch of $r(z)$ becomes realized as a mismatch of the values of the Hubble constant determined by the two methods \cite{Knox:2019rjx,Mortsell:2018mfj}. 

Despite of intense efforts \cite{Efstathiou:2020wxn,Kazantzidis:2020tko,Kazantzidis:2020xta,Sapone:2020wwz} it has not been possible to reliably identify systematic errors of the calibrators used in the context of the two methods. For example parallax data from Gaia have recently confirmed \cite{Soltis:2020gpl,Riess:2020fzl} the calibration of Cepheid stars while attempts to recalibrate the sound horizon assuming \eg  neutrino self interactions \cite{Blinov:2019gcj}, early dark energy \cite{Poulin:2018cxd,Sakstein:2019fmf,Agrawal:2019lmo,Lin:2019qug,Braglia:2020bym,Niedermann:2020dwg,Smith:2020rxx} or modified gravity \cite{Benisty:2019pxb,Ballardini:2020iws} have either failed to significantly reduce the tension level \cite{Ballardini:2020iws,DAmico:2020ods} or produced new tensions with other cosmological data \cite{Haridasu:2020pms,Krishnan:2020obg} (including growth rate from weak lensing \cite{Hildebrandt:2016iqg,Joudaki:2017zdt,Kohlinger:2017sxk,Abbott:2017wau} and peculiar velocities \cite{Macaulay:2013swa,Kazantzidis:2018rnb,Nesseris:2017vor,Skara:2019usd,Kazantzidis:2019dvk,Kazantzidis:2018jtb}). It therefore becomes increasingly probable that the mismatch in the high-z -low-z form of $r(z)$ is indeed a physical effect \cite{DiValentino:2021izs} that will require deformation of $H(z)$ from its \plcdm form and/or some other modification of late time physics.

Attempts to consider smooth deformations of $H(z)$ \cite{Alestas:2020mvb,DiValentino:2016hlg,Smith:2019ihp,Vagnozzi:2019ezj,Li:2019yem,DiValentino:2020naf,Krishnan:2020vaf} at $z\simeq O(1)$ have been successful in matching $r(z_{rec})$ with $r(z=0)$ but have been unable to match the value of $r(z\simeq O(1))$ which is strongly constrained by BAO and SnIa data to be close to the form indicated by \plcdmnospace. 

A remaining possibility is that of an abrupt deformation of $H(z)$ at $z\lesssim 0.1$ ($H(z)$ transition). Such a deformation has been considered in previous studies \cite{Mortonson:2009qq,Benevento:2020fev,Dhawan:2020xmp} as a discontinuity of $H(z)$ occurring at $z_t<0.1$. It was shown however, that if such a feature occurs at $z_t<0.01$ \ie below the redshift where Hubble flow starts, it would be undetectable by standard candles \cite{Benevento:2020fev} and thus it would not be able to justify the  measured decreased value of $r(z)$ at low $z$. On the other hand, if it occurred at $0.01<z_t<0.1$ with the proper amplitude to reduce $r(z)$ to the required level, it would have to produce a step-like feature in the SnIa Hubble diagram with amplitude $\Delta m=0.2$. A discontinuity with such an amplitude is inconsistent with the Pantheon data. It is therefore clear that even though the existence of a feature in the form of $H(z)$ at $z_t<0.1$ is likely, this feature is severely constrained by both the constraints on the comoving distance $r(z)$ at $z>0.1$ as well as by the measured SnIa magnitudes which are not consistent with a large step-like discontinuity.
This problem can be avoided by assuming a transition of the SnIa absolute magnitude $M$ at $z_t\in [0.01,0.1]$ which can nullify the required step-like feature of the apparent magnitudes while being consistent with value of the absolute magnitude implied by local Cepheid calibrators. In the present analysis we propose such a feature in the form of a a transition of the SnIa absolute magnitude accompanied by transition of the equation dark energy of state parameter $w(z)\equiv\frac{p_{de}(z)}{\rho_{de}(z)}$ \cite{Keeley:2019esp,Bassett:2002qu}.  

In particular, we consider a transition of $w(z)$ as
\be
w(z)=-1+\Delta w\;\Theta (z_t-z)
\label{wanz}
\ee
while also allowing for a corresponding transition for the SnIa absolute magnitude $M$ (due to fundamental physics changes that accompany the $w$ transition) of the form
\be
M(z)=M_C+\Delta M \;\Theta (z-z_t)
\label{manz}
\ee
where $\Theta$ is the Heaviside step function, $M_C=-19.24$ is the SnIa absolute magnitude calibrated by Cepheids \cite{Camarena:2021jlr,Camarena:2019moy} at $z<0.01$ and $\Delta M$, $\Delta w$ are parameters to be fit by the data. 

The equation of state parameter determines the gravitational properties and the evolution of dark energy density $\rho_{de}$. From energy momentum conservation $d(\rho_{de} a^3)=-p_{de} d(a^3)$ it is easy to show that the evolution of dark energy density is obtained as
\begin{widetext}
\be
\rho_{de}(z)=\rho_{de}(z_p)\int_{z_p}^z\frac{dz'}{1+z'}(1+w(z'))=\rho_{de}(z_p)\left(\frac{1+z}{1+z_p}\right)^{3(1+w)}
\label{rhodeevol}
\ee
\end{widetext}
where in the last equality a constant $w$ was assumed and $z_p$ is a pivot redshift which may be assumed equal to the present time or equal to the transition time $z_t$.
Eqs. (\ref{wanz}) and (\ref{rhodeevol}) imply a continuous Hubble expansion rate $h(z)\equiv H(z)/100km/(sec\cdot Mpc)$ of the form
\begin{widetext}
%\ba
%h(z)^2 & =& \omega_m (1+z)^3 +\omega_r (1+z)^4+(h^2-\omega_m-\omega_r) %\left(\frac{1+z}{1+z_t}\right)^{3\; \Delta w} \; \;  z<z_t \label{hzlwpt1} \\
%h(z)^2&=& \omega_m (1+z)^3 +\omega_r (1+z)^4+(h^2-\omega_m-\omega_r)  \; \;  \; \; \;  % \; \;  \; z>z_t  
%\ea

\begin{equation}
\begin{aligned}
    h_w(z)^2 & \equiv \omega_m (1+z)^3 +\omega_r (1+z)^4+(h^2-\omega_m-\omega_r) \left(\frac{1+z}{1+z_t}\right)^{3\; \Delta w}   &z<z_t  \\
h_w(z)^2&\equiv \omega_m (1+z)^3 +\omega_r (1+z)^4+(h^2-\omega_m-\omega_r)   &z>z_t  
\end{aligned}
\label{hzlwpt}
\end{equation}

\end{widetext}
where  $\omega_m\equiv \Omega_{0m} h^2$, $\omega_r\equiv \Omega_{0r} h^2$ are the matter  and radiation density parameters assumed fixed to their \plcdm values in the next section and $h$ is a parameter distinct from the rescaled measurable Hubble parameter $h_w(z=0)$\footnote{The parameter $h$ would be equal to the measured rescaled Hubble parameter $h_w(z=0)$ in the limit $z_t\rightarrow 0$.}. In what follows we assume $0.01<z_t<0.1$ and define $h_{local}\equiv 0.74$ and $h_{CMB}\equiv 0.674$  which correspond to the Hubble constant values obtained with local standard candle measurements of $r(z)$ ($H_0=H_0^{R19}$)  and sound horizon standard ruler measurements ($H_0=H_0^{P18}$ calibrated by \plcdmnospace) respectively.

In the context of the above Late $w-M$ Phantom Transition ($LwMPT$) model the following interesting questions emerge:
\begin{itemize}
    \item What is the functional form of $\Delta w(z_t)$ so that $h_w(z=0)=h_{local}$ as implied by local measurements while maintaining the required \plcdm form of $r(z)$ for $z\gg z_t$?
    \item
    How closely does the $LwMPT$ model reproduce the form of the \plcdm comoving distance $r(z)$ for $z>z_t$? How does this form of $r(z)$ compare with the corresponding form of the $H(z)$ transition?
    \item How does the quality of fit of the $LwMPT$ model to cosmological data (CMB, SnIa, BAO and SH0ES) compare with the corresponding quality of fit of typical models that utilize smooth deformations of $H(z)$ to address the $H_0$ tension?
    \item What are the favored values of $\Delta w$, $\Delta M$ and what are the implications for general relativity and for the future evolution of the universe?
\end{itemize}

In the present analysis we address the above questions. The structure of this paper is the following: In the next section we investigate analytically the ability of the $LwMPT$ model (\ref{hzlwpt}) to reproduce the \plcdm form of the comoving distance for $z>z_t$ while keeping $h_w(z=0)=0.74$. We also identify the values $\Delta w(z_t)$ that achieve this goal using an analytical approach. In section \ref{sec:LwPTcomp} we use cosmological data (CMB, SnIa, BAO and SH0ES) to identify the best fit parameter values for various transition redshifts $z_t$ and identify the improvement of the quality of fit as $z_t$ decreases down to the minimum acceptable value $z_t\simeq 0.02$. We also compare this quality of fit to the data with the \plcdm model (without the SH0ES datapoint) and with a typical smooth $H(z)$ deformation model ($wCDM$) that is designed to address the Hubble tension. Finally in section \ref{sec:Conclusion} we summarize the main results of our analysis and discuss the implications of these results for the future evolution of the universe if this model is indeed realized in Nature. We also discuss possible future extensions of this analysis.

\section{The cosmological comoving distance in the $LwMPT$ model}

\begin{figure}[b]
\centering
\includegraphics[width = 0.47 \textwidth]{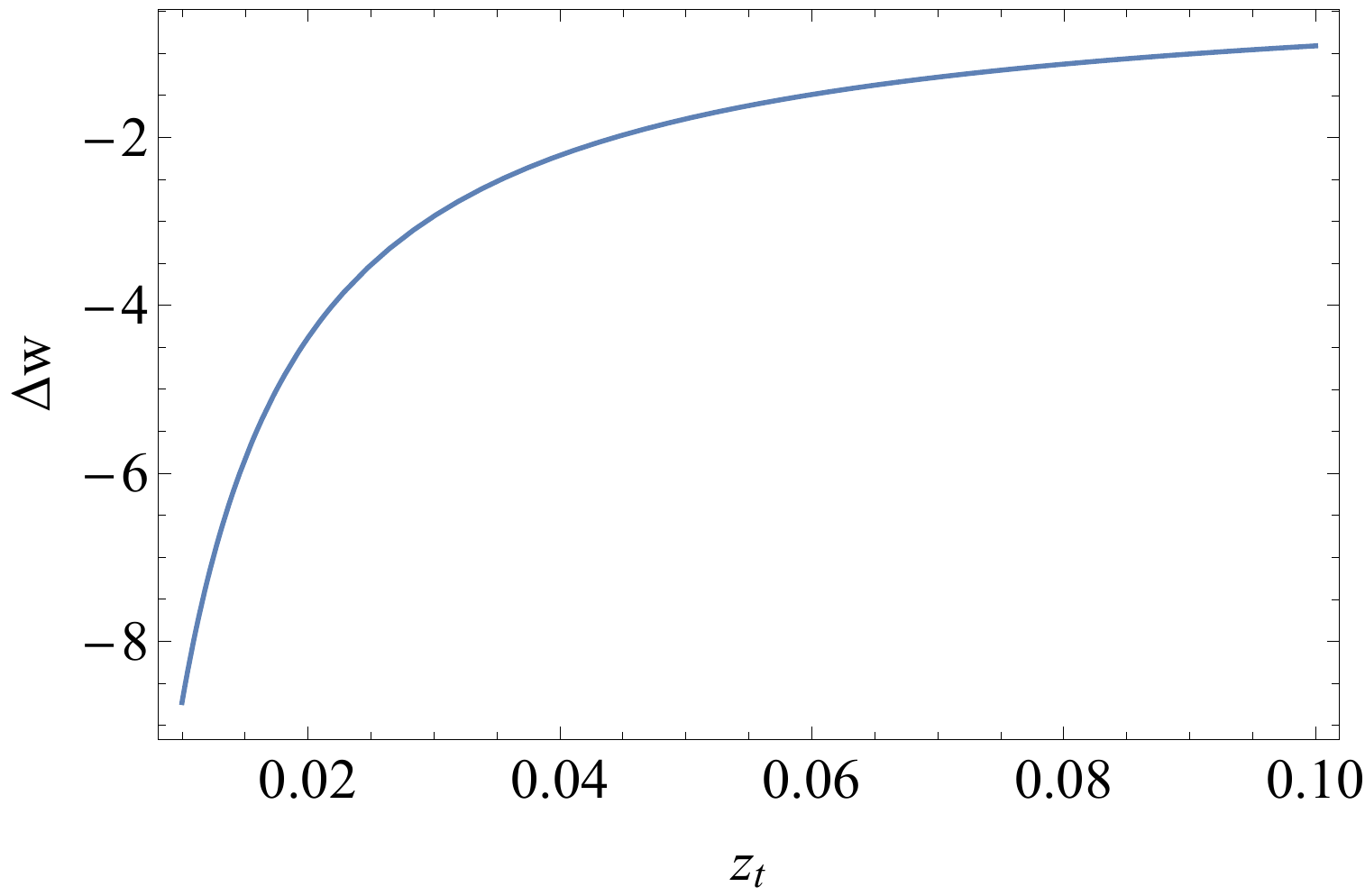}
\caption{The equation of state shift $\Delta w$ required for $h_w(z=0)=h_{local}$ as a function of the transition redshift $z_t$. Notice the strongly phantom behavior of the dark energy equation of state $w=-1-\Delta w$ for $z<z_t$.}
\label{fig1}
\end{figure} 

In order to fix the parameters $\omega_m$, $\omega_r$, $h$ and $\Delta w$ in the $LwMPT$ ansatz (\ref{hzlwpt}) we impose the following conditions:
\begin{itemize}
\item
It should reproduce the comoving distance corresponding to \plcdm $r_\Lambda$ for $z\gg z_t$ where
\be
r_\Lambda(z) \equiv \int_0^z \frac{dz'}{\omega_m (1+z')^3 +\omega_r (1+z')^4+(h^2-\omega_m-\omega_r)}
\label{rldef}
\ee
where $\omega_m\equiv \Omega_{0m} h^2=0.143$, $\omega_r\equiv \Omega_{0r} h^2 =4.64\times 10^{-5}$ and $h=h_{CMB}=0.674$.
\item
It should reproduce the local measurements of the Hubble parameter 
\be
h_w(z=0)=h_{local}=0.74.
\label{hw0cons}
\ee
\end{itemize}
The first condition fixes the parameters $\omega_m$, $\omega_r$ and $h$ to their \plcdm best fit values.
Since we consider $z_t<0.1\ll 1$ it is straightforward to obtain an upper bound for the relative difference 
\be
\frac{\Delta r}{r}(z)\equiv \frac{r_w(z)-r_\Lambda(z)}{r_\Lambda(z)}<\frac{h_{local}-h_{CMB}}{h_{CMB}}\simeq 0.1
\ee
where $r_w(z)\equiv \int_0^z \frac{dz'}{h_w(z')}$ is the comoving distance corresponding to the $LwMPT$ model (\ref{hzlwpt}). $\frac{\Delta r}{r}(z)$ is maximum at $z=0$ and decreases rapidly as $z$ increases as demonstrated below. 

The second condition imposes the constraint (\ref{hw0cons}) on eq. (\ref{hzlwpt}) and leads to a relation between $\Delta w$ and $z_t$ of the form (here we neglect $\omega_r$ as it has practically no effect on $\Delta w$)
\be
\Delta w=\frac{Log\left(h^2-\omega_m\right) - Log\left(h_{local}^2 - \omega_m\right)}{3 Log(1+z_t)}
\label{dwform}
\ee
where $h=h_{CMB}=0.674$ and $\omega_m=\Omega_{0m} h^2=0.143$ as implied by the first condition and for consistency with the CMB anisotropy spectrum.
In Fig. \ref{fig1} we show a plot of $\Delta w(z_t)$ demonstrating the strongly present day phantom behavior of dark energy implied by this class of models. 

\begin{figure*}[ht!]
\centering
\includegraphics[width = 0.9 \textwidth]{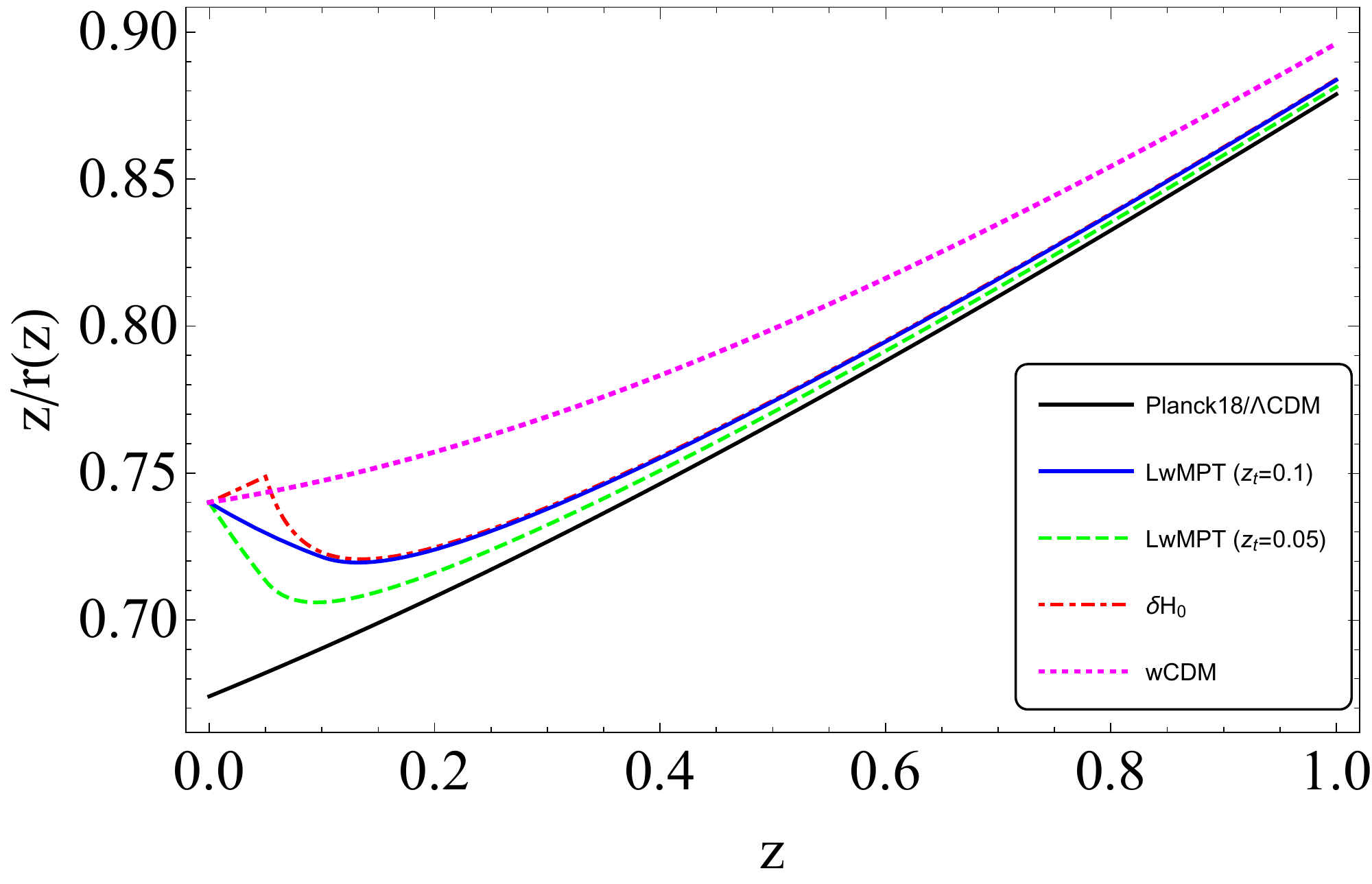}
\caption{The function $f(z)=z/r(z)$ where $r(z)$ is the comoving distance to redshift $z$ for the cosmological models \plcdm (black continuous line), $wCDM$ with $w=-1.2$ (magenta dotted line), $H(z)$ transition (\ref{hptdef}) with $z_t=0.05$ and  $\frac{\delta h}{h}=(h_{local}-h_{CMB})/h_{CMB}$ (red dot-dashed line), $LwMPT$ with $z_t=0.05$ and $w(z<z_t)=-1-\Delta w=-2.78$ as indicated by eq. \eqref{dwform} (green dashed line) and $LwMPT$ with $z_t=0.1$ and $w(z<z_t)=-1-\Delta w=-1.91$ as indicated by eq. \eqref{dwform} (blue continuous line). Notice that even though all three models approach $r_\Lambda(z)$ asymptotically, the two $LwMPT$ models remain closest to the \plcdm comoving distance $r_\Lambda(z)$ while at the same time they are consistent with the local measurement of the Hubble constant since $h_w(z=0)=0.74$. }
\label{fig2}
\end{figure*} 
It is of interest to compare the form of the comoving distance $r(z)$ predicted in the context of the $LwMPT$ model $r_w(z)$ with other proposed $H(z)$ deformations for the resolution of the Hubble tension. In Fig. \ref{fig2}  we show a plot of the function $f(z)\equiv z/r(z)$ (whose $z\rightarrow 0$ limit is the Hubble constant) for three proposed $H(z)$ deformation resolutions of the Hubble tension: the $LwMPT$ model, the $H(z)$ transition model and the $wCDM$ with fixed $w=-1.22$ model \cite{Alestas:2020mvb,DiValentino:2016hlg,Vagnozzi:2019ezj}. The $H(z)$ transition model is defined as
\begin{widetext}
\be
h_\delta(z)^2 \equiv (1+\frac{\delta h}{h} \;\Theta(z_t-z))^2\left[\omega_m (1+z)^3 +\omega_r (1+z)^4+(h^2-\omega_m-\omega_r)\right]
\label{hptdef}
\ee
\end{widetext}
where $\frac{\delta h}{h}=\frac{h_{local}-h_{CMB}}{h_{CMB}}$, $h=h_{CMB}$ and $\omega_m$, $\omega_r$ are assumed fixed to their \plcdm best fit values. The fixed $w$ ($wCDM$) smooth $H(z)$ deformation model is defined as
\begin{widetext}
\be
h_{wf}(z)^2\equiv \omega_m (1+z)^3 +\omega_r (1+z)^4+(h^2-\omega_m-\omega_r)(1+z)^{3(1+w)}
\label{wcdm}
\ee
\end{widetext}
where $w=-1.22$, $h=h_{local}$ and $\omega_m$, $\omega_r$ are assumed fixed to their \plcdm best fit values \cite{Alestas:2020mvb}.

All three models that address the $H_0$ tension shown in Fig. \ref{fig2} satisfy by construction two necessary conditions 
\ba
&&h(z=0)= h_{local} \\
&&r(z) \rightarrow  r_\Lambda(z) \; \;for  \; \; z\gtrsim O(1).
\ea

These conditions along with the fact that we fix the parameters $\omega_m$ and $\omega_r$ to their best fit \lcdm values secure the fact that all three models produce the same CMB anisotropy spectrum as \plcdm while at the same time they predict a Hubble parameter equal to its locally measured value $h(z=0)= h_{local}$. However, the three models do not approach the \plcdm comoving distance $r_\Lambda(z)$ with the same efficiency as $z$ increases. As is clearly seen in Fig. \ref{fig2},  the $LwMPT$ model with both $z_t=0.1$ and $z_t=0.05$ approaches $r_\Lambda(z)$ faster than the other two models. Since \plcdm provides an excellent fit to most geometric cosmological probes at $z>0.1$ it is anticipated that $LwMPT$ will produce a better fit to cosmological data than the smooth deformations of $H(z)$ like $wCDM$ or the discontinuous $H(z)$ transition model which produces an unnatural step in $r(z)$ and  moves away from $r_\Lambda(z)$ for $z<z_t$ as $z$ increases. This improved quality of fit is also demonstrated in the next section.

Any $H(z)$ deformation models that address the Hubble tension should not only be consistent with the locally measured value of the Hubble parameter $H_0$ and with the \plcdm form of $H(z)$. It should also be consistent with the value $M_C$ of the absolute magnitude of SnIa as determined by Cepheid calibrators \cite{Camarena:2021jlr,Camarena:2019moy}. This may be seen by considering the equation that connects the SnIa measured apparent magnitudes at redshift $z_i$ with the Hubble free luminosity distance and the Hubble parameter which may be written as
\be 
m(z_i) =  M  - 5 \log_{10}\left[H_0\cdot {\text{Mpc}/c}\right]+5 \log_{10}(D_L(z_i))  + 25  \label{mB2}
\ee
where $D_L(z)=H_0 \, d_L(z)/c$ is the Hubble free luminosity distance. Given the measured $m(z_i)$ datapoints the best fit Hubble parameter in the context of local measurements  can decrease to become consistent with the  sound horizon calibrator by either decreasing $D_L(z)$ (deforming $H(z)$) or by decreasing the absolute magnitude $M$. Such a decrease of $M$ can be achieved either by discovering a systematic effect of the Cepheid calibrators or by assuming an $M$ transition at $z\geq 0.01$ due to an abrupt change of fundamental physics. The deformation of $D_L(z)$ is severely constrained by the standard ruler constraints based on the sound horizon  (CMB and BAO) and even though it is most efficient in the context of very late transitions as the one discussed in the present analysis it may still not be enough to compensate with the decrease of $H_0$ while keeping $M$ fixed to its Cepheid calibrated value $M=M_C$. A common error made in late time approached of the Hubble tension is to either marginalize over $M$ with a flat prior or allow it to vary along with the cosmological parameters in the context of the maximum likelihood method. This may lead to a best fit value of $M$ that is inconsistent with the Cepheid measured value $M_C$ thus invalidating the results of such analysis. This problem may be overcome by allowing for a transition of $M$ from its measured value $M_C$ at $z<z_t\simeq 0.01$ to a value lower than $M_C$ by $\Delta M\geq  -5 \log_{10}\left[\frac{H^{\rm R19}_0}{H^{\rm P18}_0}\right]\simeq -0.2$  at $z>z_t\simeq 0.01$. Thus in the next section we allow $M$ to vary along with the cosmological parameters in order to achieve a good fit to the data and at the end we determine the required magnitude of the $M$ transition assumed to take place at the same time as the $w$ transition $z=z_t$ in the context of a common physical origin. Note however that for $\Delta M=-  5 \log_{10}\left[\frac{H^{\rm R19}_0}{H^{\rm P18}_0}\right]\simeq -0.2$ the $M$ transition may be sufficient for the resolution of the $H_0$ tension with no need for a $w$ transition.

\section{Fitting $LwMPT$ to cosmological data and comparison with $wCDM$}
\label{sec:LwPTcomp}

In this section we use a wide range of cosmological data to estimate the quality of fit and the best fit parameter values of three representative cosmological models:
\begin{itemize}
    \item The $LwMPT$ class of models defined by a Hubble expansion rate similar to that of eq. (\ref{hzlwpt}).  Here we remove the constraint  $w_>=-1$ for $z>z_t$ as well as the constraint $\omega_m=0.143$. Thus the model is now allowed to have three free parameters for each fixed value of $z_t$: $w_>$, $w_<\equiv w_>+\Delta w$ and $\omega_m$. However, as discussed below, the additional free parameters end up  constrained by the data very close to the values considered fixed in the previous section. The constraint $h(z=0)=h_{local}$ is imposed as a prior in the analysis.
    \item
    The $wCDM$ model defined in (\ref{wcdm}) with two free parameters: $w$ and $\omega_m$. The constraint $h(z=0)=h_{local}$ is imposed as a prior in the analysis.
    \item
    The \lcdm model defined by (\ref{wcdm}) with $w=-1$. No constraint for $h(z=0)$ is imposed on this model in order to maximize the quality of fit to the data and use the model as a benchmark for comparison with the other models that address the $H_0$ tension. Thus we use the term u\lcdm (``u" for ``unconstrained") to denote it. It is considered as a baseline to compute residuals of $\chi^2$ to compare the other two representative models. Its best fit parameter values ($\Omega_{0m}=0.312 \pm 0.006, H_0=67.579 \pm 0.397$) in the context of the dataset we use are almost identical with \plcdmnospace.
\end{itemize}
We use the following data to identify the quality of fit of these models
\begin{itemize}
    \item The Pantheon SnIa dataset \cite{Scolnic:2017caz} consisting of 1048 distance modulus datapoints in the redshift range $z\in[0.01,2.3]$.
    \item A compilation of 9 BAO datapoints in the redshift range $z\in[0.1,2.34]$. The compilation is shown in the Appendix.
    \item The latest \plcdm CMB distance prior data (shift parameter $R$ \cite{Elgaroy:2007bv} and the acoustic scale $l_a$ \cite{Zhai:2018vmm}). These are highly constraining datapoints based on the observation of the sound horizon standard ruler at the last scattering surface $z\simeq 1100$. The covariance matrix of these datapoints and their values are shown in the Appendix.
    \item
    A compilation of 41 Cosmic Chronometer (CC) datapoints in the redshift range $z\in[0.1,2.36]$. These datapoints are shown in the Appendix and have much less constraining power than the other data we use.
\end{itemize}

Using these data (total of 1100 datapoints) we used the maximum likelihood method \cite{Arjona:2018jhh} to minimize the total $\chi^2$ defined as 
\be 
\chi^2=\chi^2_{CMB}+\chi^2_{BAO}+\chi^2_{CC}+\chi^2_{Panth}
\ee
and calculate the residual $\Delta \chi^2$  with respect to the u\lcdm model for the $LwMPT$ class (as a function of $z_t$) and for $wCDM$. Since the CMB data are the most constraining, we have found the anticipated best fits $\omega_m\simeq 0.143$  and $w=-1.22$ for $wCDM$ (see Ref. \cite{Alestas:2020mvb} for a detailed analysis of these results). 

\begin{figure}[ht]
\centering
\includegraphics[width = 0.47 \textwidth]{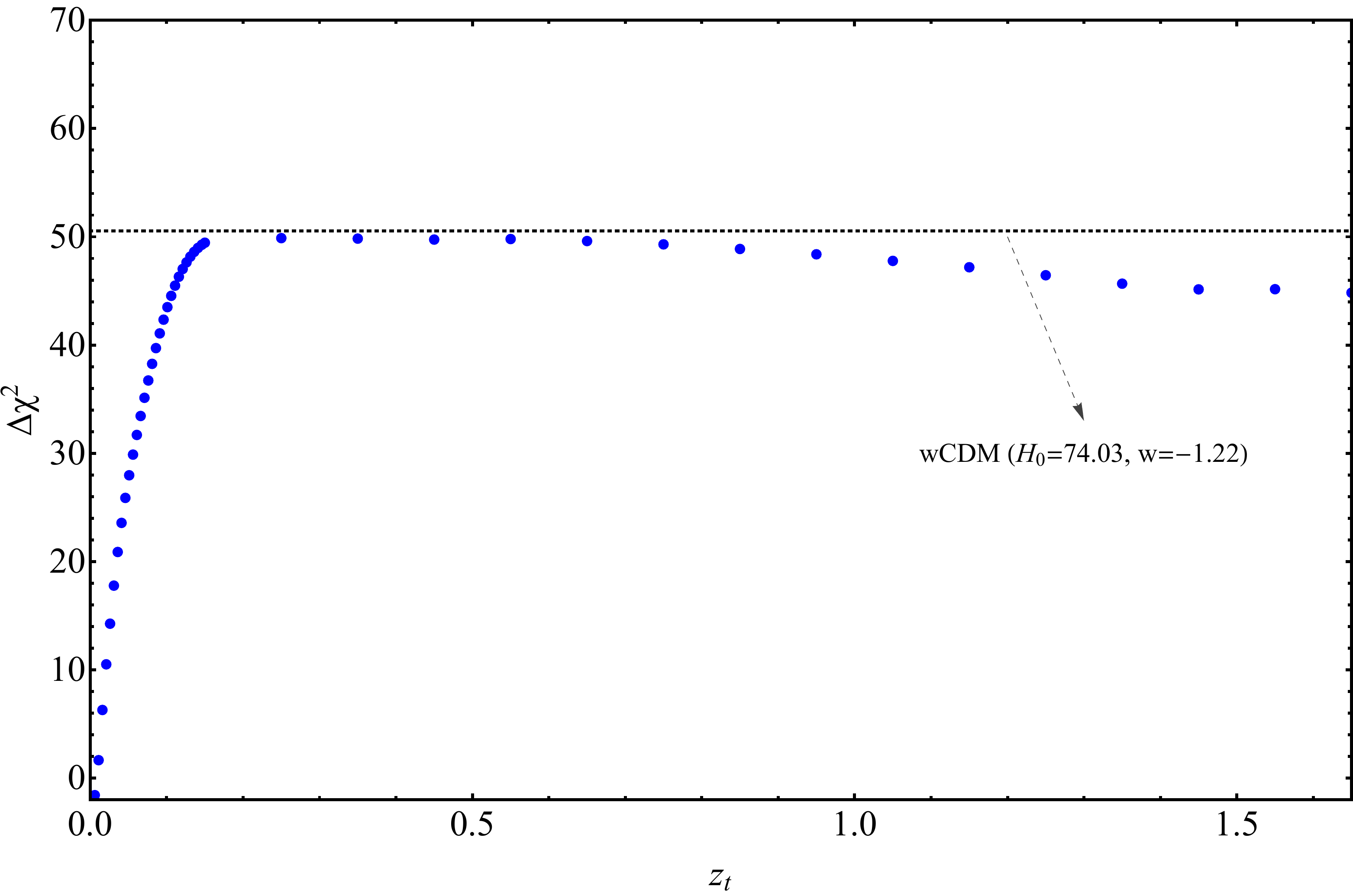}
\caption{The residuals $\Delta\chi^2$ plotted against the values of the transition redshift $z_{t}$ for the $LwMPT$ (blue dots) and $wCDM$ (black dotted line) with $w=-1.22$. The $LwMPT$ model seems to achieve a significantly better fit for small $z_{t}$ values.} 
\label{fig3}
\end{figure}

\begin{figure}[b]
\centering
\includegraphics[width = 0.47 \textwidth]{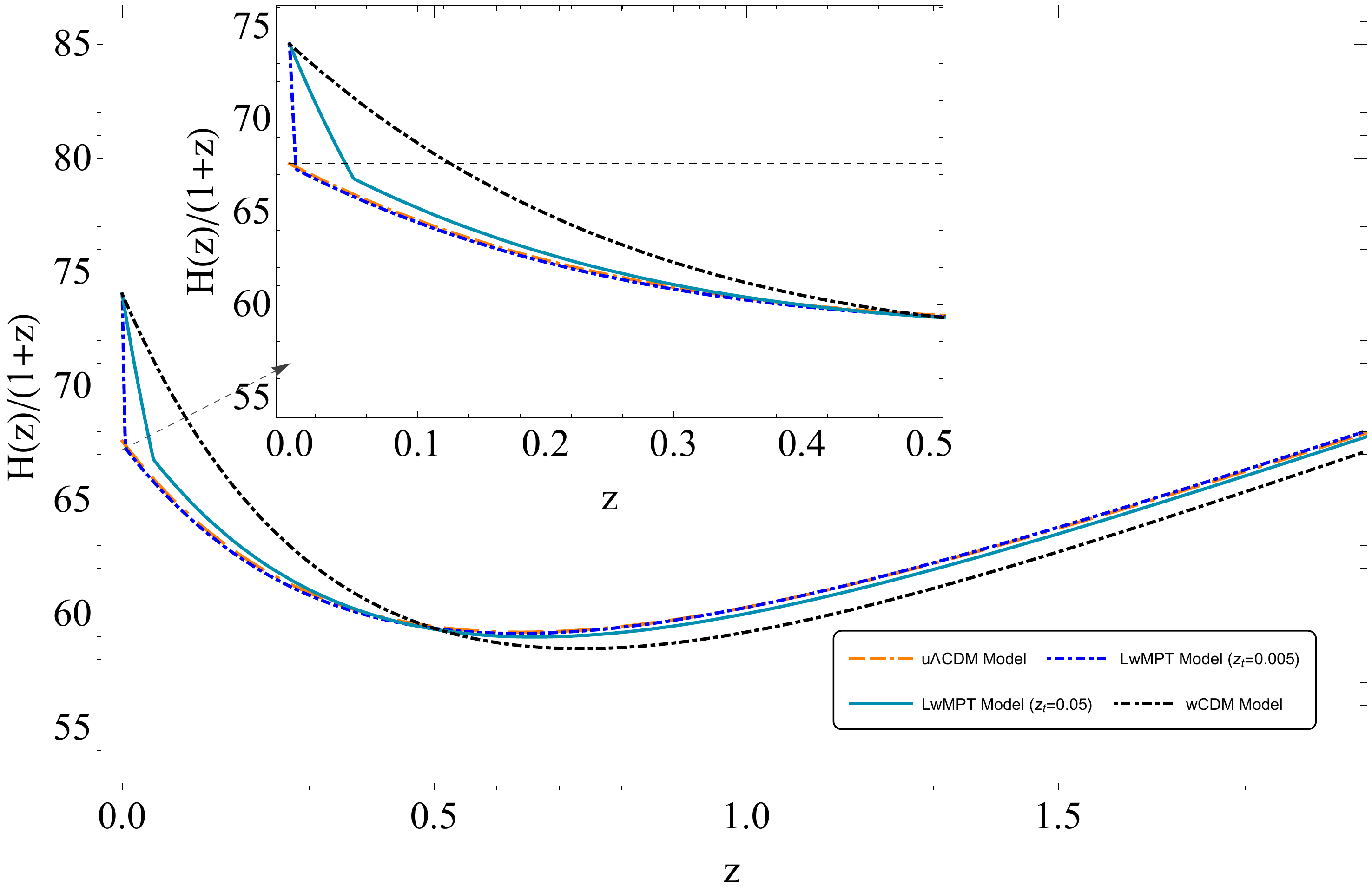}
\caption{The forms of the comoving Hubble parameter $H(z)/(1+z)$ for two $LwMPT$ models with $z_t<0.1$, the best fit $wCDM$ and u\lcdmnospace.}
\label{fig4}
\end{figure}

In Fig. \ref{fig3} we show the residuals $\Delta \chi^2$ for the best fit $LwMPT$ models as a function of $z_t$ (blue points) and the corresponding residual $\Delta \chi^2$ for the best fit $wCDM$ model (horizontal black line). As mentioned in the previous section no prior is imposed on $M$ but at the end we will identify the magnitude $\Delta M$ of the required $M$ transition. The rapid improvement of the fit compared to $wCDM$ for the $LwMPT$ models as $z_t$ decreases below $z_t\simeq 0.15$ is clear. The best fit parameter values for $w_<$ ($z<z_t$) and $w_>$ ($z>z_t$) are shown in Table \ref{tab:ztchi2w0wa}. In parenthesis next to each $w_<$ best fit we show the predicted value in the context of the analysis of the previous section (eq. (\ref{dwform})) which assumes $w_>=-1$.

\begin{figure*}[ht!]
\centering
\includegraphics[width = 1.0 \textwidth]{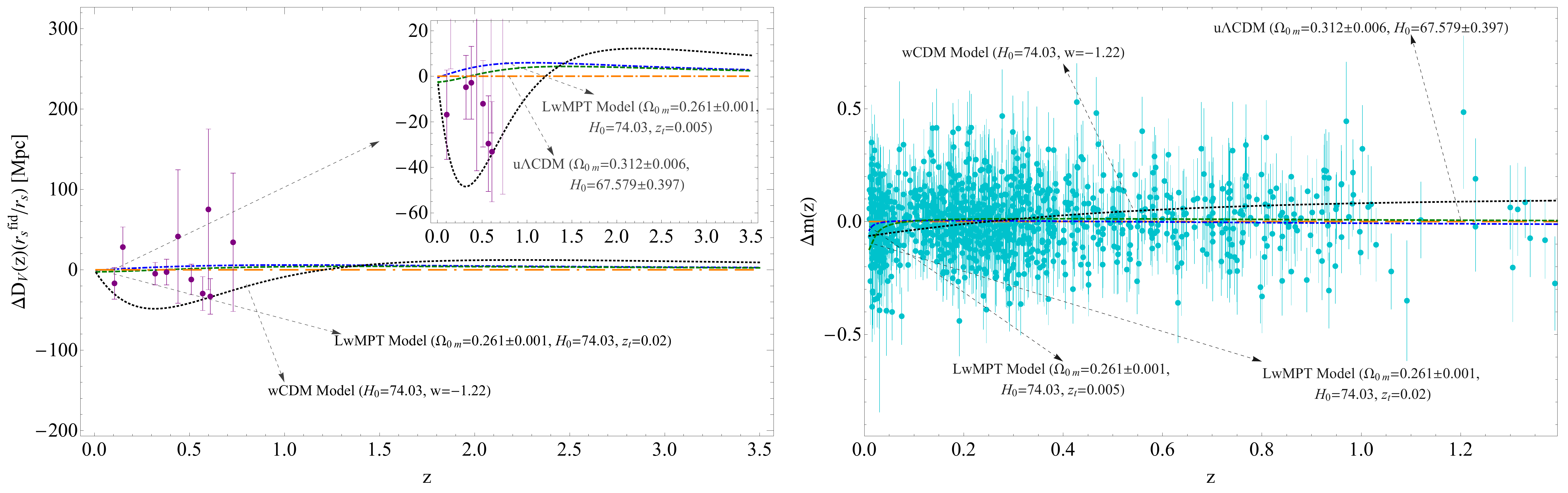}
\caption{{\it{Left panel:}} BAO data residuals $\Delta D_V\times \frac{r_s^{fid}}{r_s}$ from the best fit u\lcdm (orange dashed line) superimposed with the best fit residual curves corresponding to $wCDM$ (dotted black line), $LwMPT$ with $z_t=0.005$ (blue dot dashed line) and  $LwMPT$ with $z_t=0.02$ (green dashed line). Notice the difficulty of smooth $H(z)$ deformation of $wCDM$ to fit the data due to the constraint imposed by the local measurements of Hubble constant. {\it{Right panel:}} The Pantheon SnIa distance modulus residuals $\Delta m$ from the best fit u\lcdmnospace. The predicted distance modulus residual curves for $wCDM$ (black dashed line), the $LwMPT$  ($z_t=0.005$) (blue dot dashed line) and the $LwMPT$  ($z_t=0.02$) (green dashed line) are also shown.}
\label{fig5}
\end{figure*}

\begin{figure*}[ht!]
\centering
\includegraphics[width = 1.0 \textwidth]{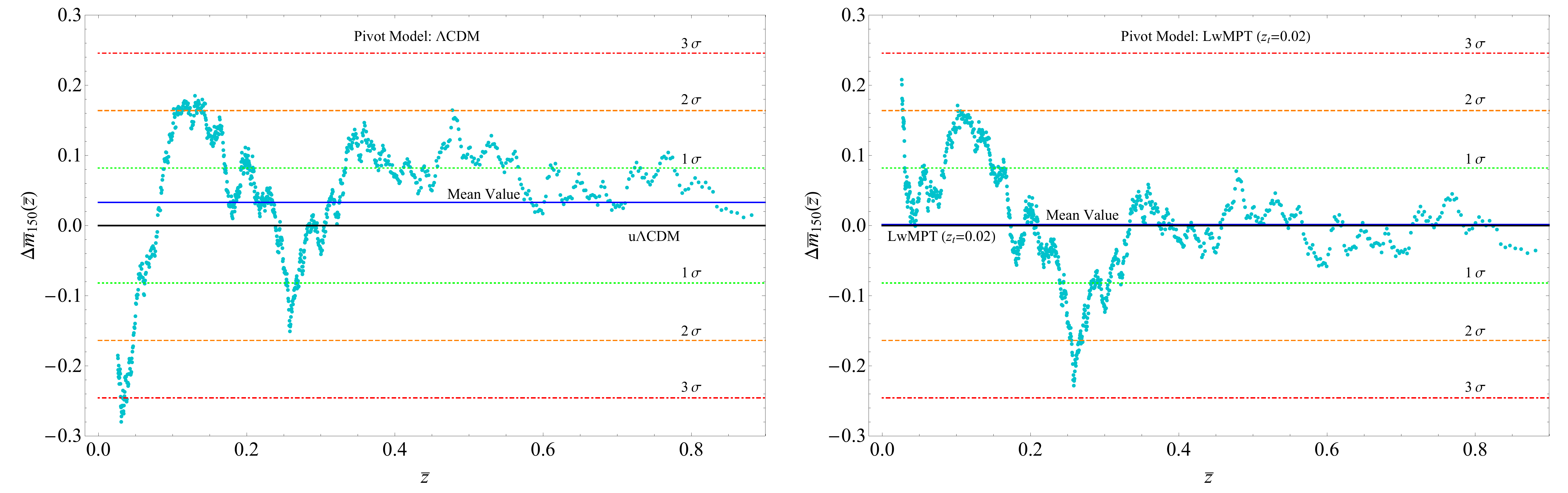}
\caption{{\it{Left panel:}} The 150 point moving average of the Pantheon SnIa standardized residual absolute magnitudes with respect to the best fit \lcdm ($\Delta{\bar m}_{150}(\bar{z})$ of eqs (\ref{movavmnj}-\ref{movavredj})). Notice the sharp and peculiar drop at $z\lesssim 0.1$ (unlikely at more than $3\sigma$ level). {\it{Right panel:}} The 150 point moving average of the Pantheon SnIa standardized residuals with respect to the best fit $LwMPT$ ($z_{t}=0.02$).  The sharp drop shown in the left panel has disappeared while the mean and the standard deviation of the moving average points have dropped significantly indicating that the $LwMPT$ is a more natural pivot model than \plcdmnospace.}
\label{fig6}
\end{figure*} 

\begin{centering}
\begin{table}[ht]
\caption{The values of the $LwMPT$ model best fit parameters $\Omega_{0m}$, $w_{<}$ ($z<z_t$) and $w_{>}$ ($z>z_t$)  corresponding to different indicative values of the transition redshift $z_{t}$, along with each case's $\Delta\chi^2$ with respect to u\lcdmnospace. In parenthesis we show the analytically predicted values of $w_<$ which were obtained from eq. (\ref{dwform}) (\ie assuming $w_>=-1$ and imposing the constraint $h(z=0)=h_{local}$ on the $LwMPT$ ansatz (\ref{hzlwpt})). Notice that the best fit values of $\Omega_{0m}$ are consistent with the CMB spectrum requirement of $\omega_m=0.143$ in view of the constraint $h(z=0)=h_{local}$ imposed in all cases.}
\label{tab:ztchi2w0wa}
\begin{tabular}{|c|c|c|c|c|c|}
\hline 
 \rule{0pt}{3ex}  
$z_t$  & $\Delta\chi^2$  & $\Omega_{0m}$ & $w_{<}$ ($z<z_t$) & $w_{>}$ ($z>z_t$) \\
    \hline
    \rule{0pt}{3ex}  
 0.005 & -1.9 & 0.2609 & $-18.44 \, (-18.4)$ & -1.005 \\
 0.01 & 0.8 & 0.2608 & $-9.93 \, (-9.7)$ & -1.001 \\
 0.02 & 9.7 & 0.2607 & $-5.28 \, (-5.3)$ & -1.011 \\
 0.04 & 23.1 & 0.2606 & $-2.93 \, (-3.2)$ & -1.037 \\
 0.05 & 27.6 & 0.2607 & $-2.48 \, (-2.8)$ & -1.049 \\
 0.06 & 31.3 & 0.2607 & $-2.19 \, (-2.5)$ & -1.059 \\
 0.08 & 37.9 & 0.2608 & $-1.81 \, (-2.1)$ & -1.085 \\
 0.1 & 43.3 & 0.2611 & $-1.58 \, (-1.9)$ & -1.115 \\
 0.2 & 50.1 & 0.2622  & $-1.22 \, (-1.4)$& -1.230 \\
 \hline
\end{tabular}
\end{table}
\end{centering}

The forms of the comoving Hubble parameter $H(z)/(1+z)$ for two $LwMPT$ models with $z_t<0.1$, the best fit $wCDM$ and u\lcdm are shown in Fig. \ref{fig4}. This figure demonstrates the efficiency of $LwMPT$ in mimicking the best fit  u\lcdm model (which is almost identical with \plcdmnospace) while at the same time addressing the Hubble tension by reaching $h(z=0)=h_{local}$ in a continuous manner. On the other hand the smoother approach of $wCDM$ is much less efficient in mimicking \plcdm and the price it pays for this inability is a much worse quality of fit compared to $LwMPT$ as shown in Fig. \ref{fig3}.

\begin{figure}[ht!]
\centering
\includegraphics[width = 0.47 \textwidth]{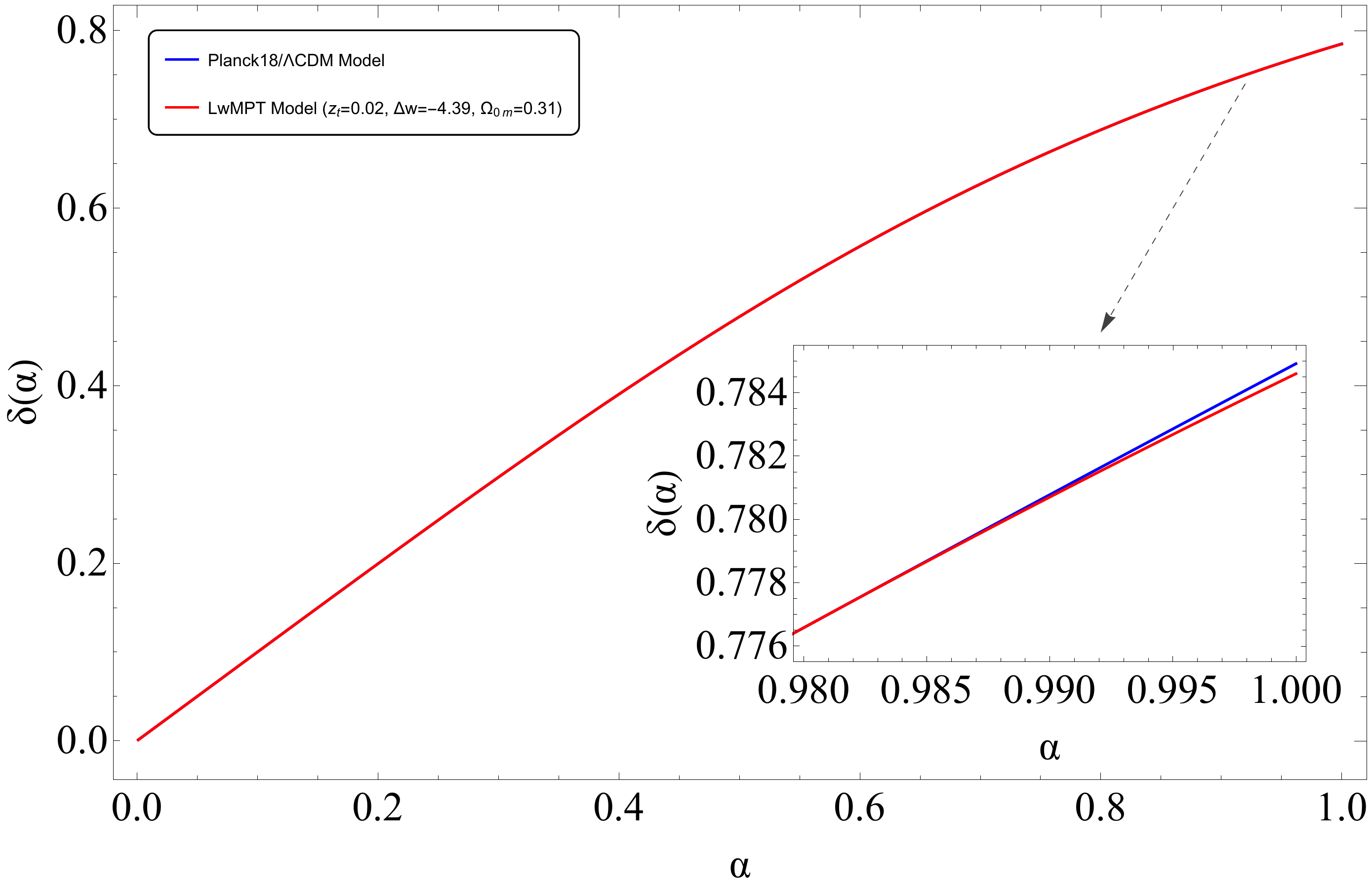}
\caption{The growth factors $\delta(\alpha)$ of the linear perturbations, for both the $LwMPT$, with $z_t = 0.02$ and $\Delta w=-4.39$ (red line), and Planck18/$\Lambda$CDM (blue line) models. Clearly, the effect of the $w$ transition on the growth factor is negligible at it occurs at $a_t\simeq 0.98$.}
\label{figdaplot}
\end{figure}

The difficulty of the smooth $H(z)$ deformation models that address the Hubble tension in fitting the BAO and SnIa data is also demonstrated in Fig. \ref{fig5} where we show the BAO and SnIa data (residuals from the best fit u\lcdmnospace) along with the best fit residuals for the $wCDM$ and $LwMPT$ models.

The right panel of Fig. \ref{fig5} indicates that the $LwMPT$ model with $z_t=0.02$ which can resolve the Hubble tension, closely mimics the apparent magnitudes of u\lcdm for  $z>z_t$ but for $z<z_t$ it predicts a small reduction of the residual apparent magnitudes. The question therefore to address is the following: {\it Is there a hint for such a statistically significant reduction of the measured absolute magnitudes in redshifts close to the transition redshift $z_t\simeq 0.02$?} Interestingly, this is indeed the case! 

The left panel of Fig. \ref{fig6} shows the $N=150$ point moving average of the standardized residual absolute magnitudes with respect to the best fit \lcdm model.  The \lcdm standardized residual apparent magnitudes are defined as
\be
\Delta \bar{m}(z_i)\equiv \frac{m_{obs}(z_i)-m_{th\Lambda CDM}(z_i,M_{bf},\Omega_{0mbf})}{\sigma_{i-tot}}
\label{barmdef}
\ee
where $\sigma_{i-tot}$ is the total error (statistical+systematic), $M_{bf}=-19.23$ and $\Omega_{0mbf}=0.30$ are the best fit parameter values of \lcdm in the context of the Pantheon data and $m_{th\Lambda CDM}$ are the corresponding theoretically predicted apparent magnitudes. The $N$ point moving average corresponding to the residual standardized datapoint point $j$ ($j\in [1,1048-N]$) is defined as
\be
\Delta{\bar m}_N^j({\bar z})\equiv \frac{1}{N}\sum_{i=j}^{j+N} \Delta{\bar m}(z_i)
\label{movavmnj}
\ee
and the corresponding redshift is 
\be
{\bar z}_N^j\equiv \frac{1}{N}\sum_{i=j}^{j+N} z_i
\label{movavredj}
\ee
For $N=150$ the left panel of Fig. \ref{fig6} shows the form of $\Delta{\bar m}_N({\bar z})$. Since the points are standardized and ignoring their correlations, we expect that the $1\sigma$ region will approximately correspond to $\sigma \simeq 1/\sqrt{N}\simeq 0.08$ which is also indicated in Fig. \ref{fig6} up to the $3\sigma$ level. Interesting features of the binned Pantheon data have been identified in previous studies \cite{Kazantzidis:2020xta, Kazantzidis:2020tko}. Related to such features is a clear abrupt drop of the moving average of the standardized residuals from the $+2\sigma$ region to the $-3\sigma$ region and beyond clearly seen in the left panel of Fig. \ref{fig6}. The deepest part of this drop is at a redshift of about $0.02$. This is precisely the type of signature anticipated in the context of the $LwMPT$ model. Once we consider the residuals with respect not to the best fit \lcdm but to the best fit $LwMPT$ model with $z_t=0.02$, this peculiar feature disappears (Fig. \ref{fig6} right panel). In addition, the standard deviation of the points of the  moving average of residuals decreases by about $20\%$ (from $0.1$ to $0.8$) while their mean value shown in Fig. \ref{fig6} drops sharply from $0.03$ to $0.001$. This is also a hint that the best fit $LwMPT$ with $z_t=0.02$ is a more natural pivot model than the best fit \lcdmnospace. This observation supports the consideration of a combined $w-M$ transition for the resolution of the Hubble tension instead of using simply an $M$ transition.

In contrast to smooth $H(z)$ deformations that in general tend to worsen the growth tension by increasing the growth rate of cosmological perturbations at early times \cite{Alestas:2021xes} the proposed ultra-late $w$ transitions have negligible effect on the growth rate of cosmological perturbations.  At $z_t=0.02$ most structures have already gone nonlinear during the $w=-1$ era and have decoupled from the effects of the background expansion. Even those fluctuations that are still linear do not have the time to respond to the change of $w$ since it occurs at very low $z$ ($z\simeq 0.02$). In addition, the emerging strongly phantom background could only lead to a suppression of the growth due the super accelerating expansion which prevents the growth of perturbations. We demonstrate this minor suppressing effect on the growth in Fig. \ref{figdaplot}, where we have solved numerically the equation for the growth of linear perturbations for the $LwMPT$ model for $z_t=0.02$ and for the required $\Delta w= -4.39$  showing that the effect on the growth factor is negligible compared to the \plcdm growth factor. If the effect of a possible gravitational transition inducing the change of $M$ were to be taken into account, the decrease of the growth factor may be shown to be large enough to resolve also the growth tension \cite{Marra:2021fvf}.

\begin{figure}[t]
\centering
\includegraphics[width = 0.47 \textwidth]{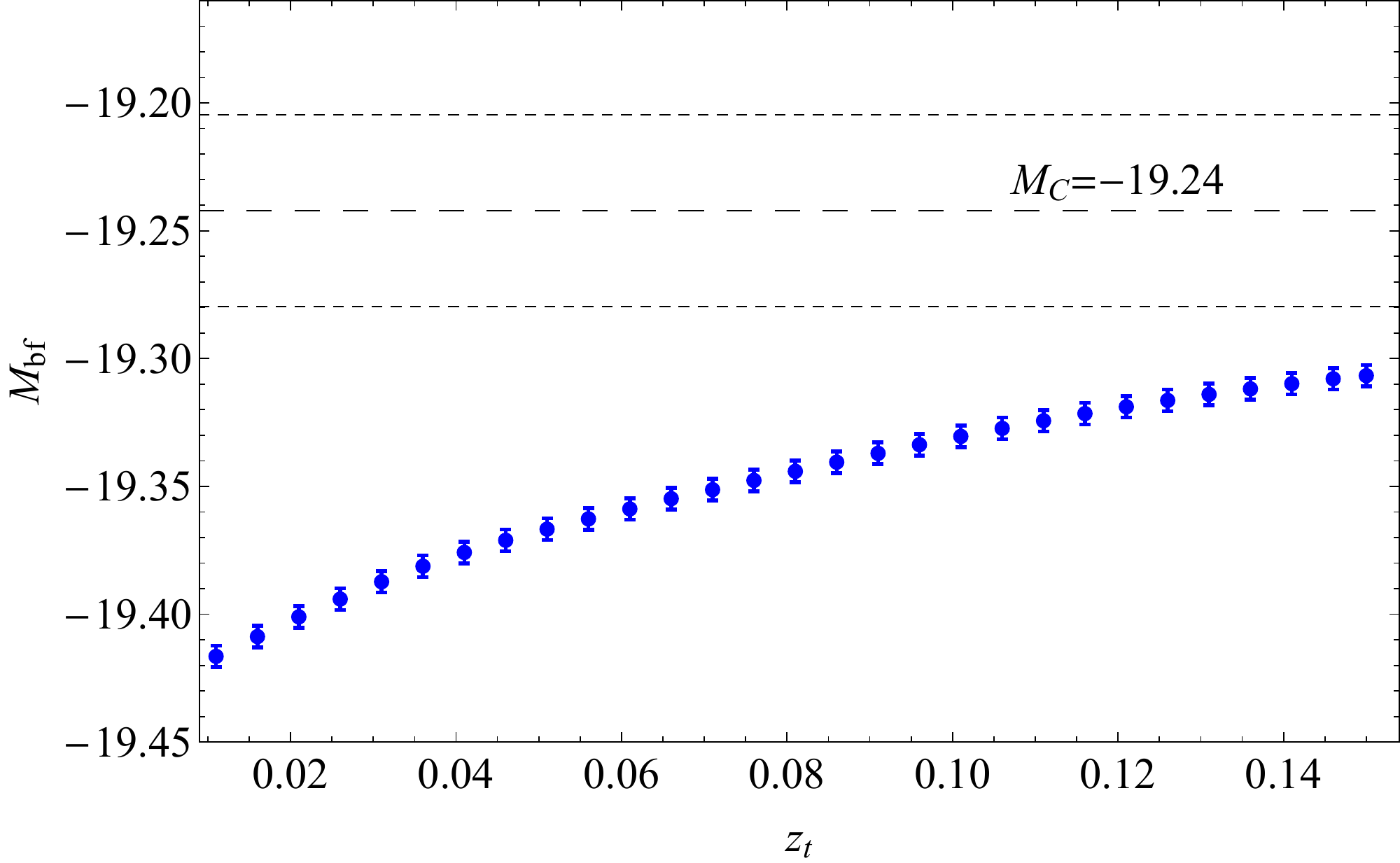}
\caption{The best fit absolute magnitude $M_{bf}$ (blue points) for various transitions $z_t$ for the $LwMPT$ model. The dashed line corresponds to the $M_C$ value indicated by Refs. \cite{Camarena:2019moy,Camarena:2021jlr}, while the dot dashed lines correspond its $1\sigma$ error. Notice that if $M_C$ is considered to be constant, the majority of the best fit values of $M_{bf}$ are more than $2\sigma$ away from the $M_C$ value. This difference reduces as $z_t$ increases.}
\label{figMplt}
\end{figure}

\begin{figure*}[ht!]
\centering
\includegraphics[width = 1.0 \textwidth]{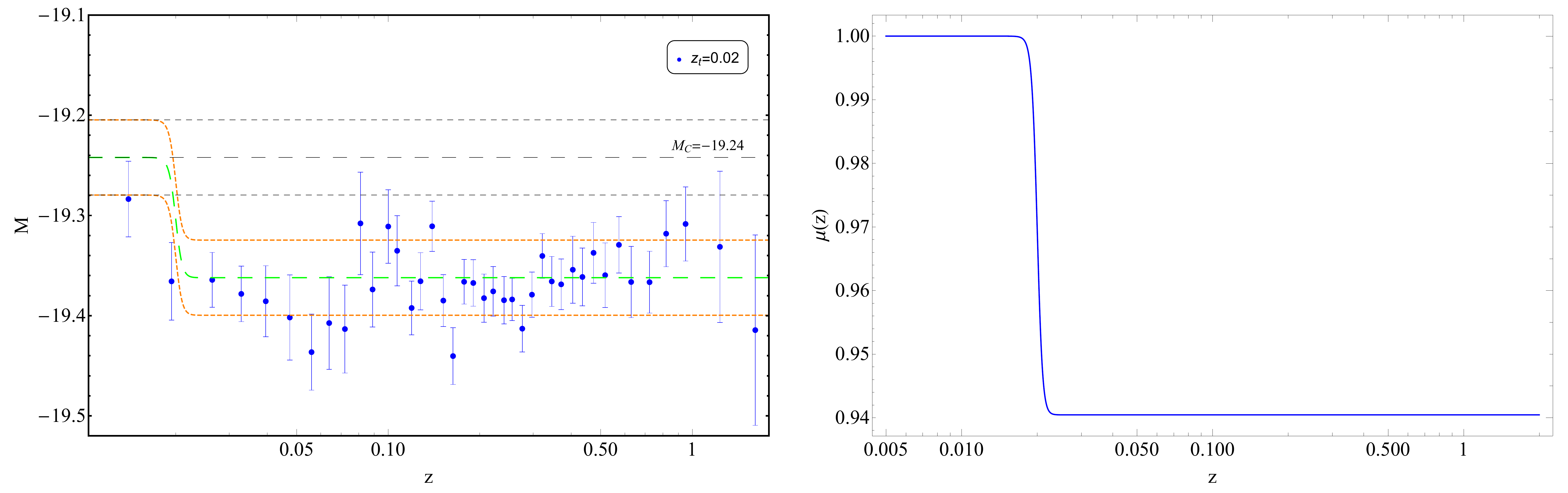}
\caption{{\it{Left panel:}} The absolute magnitude $M$ as a function of redshift $z$. The straight dashed line corresponds to the fixed $M_C$ value \cite{Camarena:2019moy,Camarena:2021jlr} from local Cepheid calibrators of SnIa  while the dot dashed lines correspond to its $1\sigma$ error.  Clearly Pantheon binned absolute magnitudes $M_i$ (blue points) corresponding to the best fit $LwMPT$ model $(z_t=0.02)$ are approximately $2 \sigma$ away from $M_C$. However, in the context of an abrupt transition of $M$ with $\Delta M\simeq -0.1$  at $z_t=0.02$, the inconsistency disappears. {\it{Right panel:}} The form of $\mu=G_{\rm eff}/G_{\rm N}$ required to induce he $M$ transition shown on the left panel.  Clearly, for $z>0.02$ $G_{\rm eff}<1$ hinting towards weaker gravity \cite{Gannouji:2018ncm,Gannouji:2020ylf,Amendola:2017orw} as indicated by other studies discussing the growth tension \cite{Hildebrandt:2016iqg,Joudaki:2017zdt,Abbott:2017wau,Kazantzidis:2019dvk, Skara:2019usd, Kazantzidis:2020xta}.}
\label{fig7}
\end{figure*}

In order to identify the magnitude of the required $M$ transition we evaluate the best fit value of the absolute magnitude $M_{bf}(z_t)$ for $z_t\it[0.01,0.15]$  (Fig. \ref{figMplt}). Notice that $\Delta M\equiv M_C-M_{bf}$ is maximum at $z_t=0.01$ and approaches the $1\sigma$ distance from $M_C$ at $z_t>0.15$. For such high values of $z_t$ however the BAO data are poorly fit and the value of $\chi^2$ increases to the level of $wCDM$. The type of the required $M$ transition for $z_t=0.02$ is shown in the left panel of Fig. \ref{fig7} where we also show the absolute magnitudes of the binned Pantheon datapoints obtained from eq. (\ref{mB2}) by solving with respect to $M$ for each datapoint and using the best fit form of $D_L(z)$ for $z_t=0.02$. Clearly, the derived absolute magnitudes are not consistent with the Cepheid calibrated value of $M_C$ but in the context of an $M$ transition with $\Delta M\simeq - 0.1$ the inconsistency disappears. The right panel of Fig. \ref{fig7} shows the required evolution of an effective Newton's constant that is required to produce the $M$ transition obtained under the assumption that the SnIa absolute luminosity is proportional to the Chandrasekhar mass which varies as $L\sim G_{\rm eff}^{b}$ with $b=-3/2$.\footnote{If $b\neq -3/2$ and especially if $b>0$ as indicated in \cite{Wright:2017rsu} under a wide range of assumptions,  then the ability of the  $LwMPT$ model to resolve the growth tension could be negatively affected.} This assumption leads to the variation of the SnIa absolute magnitude $M$ with $\mu\equiv \frac{G_{\rm eff}}{G_{\rm N}}$ ($G_{\rm N}$ is the locally measured Newton's constant) as \cite{Amendola:1999vu,Gaztanaga:2001fh,Kazantzidis:2019dvk}
\be 
\Delta M=\frac{15}{4} \, log_{10} \, \left( \mu \right)
\ee
which implies that for $\Delta M\simeq -0.1$ we have a $6\%$ reduction of $\mu$.  

Notice that if the SnIa data analysis assumes a fixed value of $M=M_C$ then the existing $m(z_i)$ data lead to a value of $H_0=74km/(sec\cdot Mpc)$ while if the transitions (\ref{wanz}) and (\ref{manz}) are assumed with $z_t=0.02$, $\Delta w=-4.3$ and $\Delta M=-0.1$ then the data analysis would lead to a value $H_0=67.5 km/(sec\cdot Mpc)$ (consistent with CMB-BAO calibration) while the true value of $H_0$ would be $H_0=74km/(sec\cdot Mpc)$ due to the $H_0$ prior imposed on the $w$ transition $\Delta w$.

\section{Conclusion-Discussion-Outlook}
\label{sec:Conclusion}

We have demonstrated using both an analytical approach and a fit to cosmological data that a Late dark energy equation of state $w$ Phantom Transition ($LwMPT$) from $w_>=-1$ ($z>z_t$) to $w_< <-1$ ($z<z_t$) at transition redshift $z_t\in[0.01,0.1]$ can lead to a resolution of the Hubble tension in a more efficient manner than smooth deformations of the Hubble tension and other types of late time transitions (the Hubble expansion rate transition). The required type of transition is a phantom transition with $w_<(z_t) \in [-2,-10]$ for $z<z_t$. The moving average statistic of the standardized residual Pantheon absolute magnitude SnIa data indicates the presence of a peculiar feature at $z<0.1$ which is consistent with the anticipated signatures of the $LwMPT$ model. Such a transition leads in general to a best fit value of the SnIa absolute magnitude that is not consistent with the value implied by local Cepheid calibrators of SnIa  \cite{Camarena:2021jlr}. Therefore late time transitions can only constitute successful resolutions of the Hubble tension if they are accompanied by a transition of the SnIa absolute magnitude due to evolving fundamental constants. We have shown that a transition of the effective gravitational constant to a value lower by about $6\%$ is sufficient to induce the required $\Delta M$ transition. This weakening of gravity may also justify the observed reduced growth of perturbations which is supported by Weak Lensing \cite{Hildebrandt:2016iqg,Joudaki:2017zdt,Abbott:2017wau} and Redshift Space Distortion data \cite{Kazantzidis:2019dvk, Skara:2019usd, Kazantzidis:2020xta} (growth tension). Therefore, this model simultaneously addresses both the Hubble and the growth tensions.

Another basic advantage of such a late time model that can fully resolve the Hubble tension is that it can fit the local distance data (BAO and SnIa) in a very effective manner. This is due to the fact that by construction it has the same quality of fit to the BAO, SnIa and CMB data as \plcdmnospace, in contrast to the usual late time smooth deformations of $H(z)$.

Moreover, there is a physical theoretical basis of the model since  it  can  be  realized  in  the  context of  modified gravity theory with a rapid gravitational transition. The rapid nature of the transition is a generic feature and can be made consistent with solar system tests with no need for screening as in other modified theories. Such models include the following:
\begin{itemize}
\item The most natural model that can induce a $LwMPT$ involves a non-minimally coupled phantom scalar field initially frozen at $\phi=\phi_0$ due to cosmic friction close to the zero point of its potential which could be assumed to be of the form $V(\phi)= s\; \phi^n$.  Such a field would initially have a dark energy equation of state $w=-1$ mimicking a cosmological constant. Once Hubble friction becomes smaller than the field dynamical (mass) scale, the field becomes free to roll up its potential (phantom fields move up their potential in contrast to quintessence fields \cite{Perivolaropoulos:2004yr,Nesseris:2006er}) and develops a rapidly changing equation of state parameter $w<-1$ and shifted $G_{\rm eff}$. Thus the universe enters a ghost instability phase which will end in a Big Rip singularity in less than a Hubble time. Such a scenario for the simple (but also generic) case of linear potential ($n=1$) has been investigated in Ref. \cite{Perivolaropoulos:2004yr}. For a general phantom potential we anticipate a redshift dependence of the equation of state $w_<=w_<(z)$ after the transition ($z<z_t$). In fact the phantom field potential could be reconstructed by demanding a form of $w_<(z)$ that further optimizes the quality of fit to the low $z$ data or by simply demanding that $w_<$ is constant.
\item A scalar-tensor modified gravity theory field  initially frozen due to Hubble friction, mimicking general relativity and a cosmological constant. Once Hubble friction becomes smaller than the field mass scale, the field becomes free to roll down its potential inducing deviations from general relativity on cosmological scales and a phantom departure from the cosmological constant. Note that scalar tensor theories can induce phantom behavior without instabilities in contrast to a simple minimally coupled scalar field \cite{Perivolaropoulos:2005yv}.
\end{itemize}
The detailed investigation of the above described dynamical scalar field evolution that can reproduce the $LwMPT$ is an interesting extension of the present analysis.

If the phantom $LwMPT$ is realized in Nature it would imply the existence of a rapidly approaching Big Rip singularity \cite{Caldwell:2003vq,Nesseris:2004uj} which may be avoided due to quantum effects \cite{Elizalde:2004mq}. Given the value of $w_<$ which emerges at approximately the present time $t_0$, it is straightforward to calculate the time $t_*$ of the Big Rip singularity assuming  that $w=w_<<-1$ at the present time $t_0$. The result is \cite{Nesseris:2004uj}
\be
\frac{t_*}{t_0}=\frac{w_<}{1+w_<}
\label{bigriptime}
\ee
 For example for $z_t=0.02$ we have $w_<\simeq -5$ which implies that the universe will end in a Big Rip singularity in less than $3.5$ billion years (for $t_0=13.8\times 10^9 yrs$).  This implies that there may be observational effects of such coming singularity on the largest bound systems like the Virgo cluster,  the Coma Cluster or the Virgo supercluster. A detailed investigation of the observational effects on bound systems of the $LwMPT$ is an interesting extension of the present analysis.

The detailed comparison of the quality of fit of the $LwMPT$ (or similar) models with a variety of smooth $H(z)$ deformation models addressing the Hubble tension would also be a useful extension. The use of full CMB spectrum data and possibly other cosmological data sensitive to the dynamics of galaxies in clusters and superclusters could also be included.

The late time sudden deformation of the luminosity distance $D_L(z)$ induced through the $w$ transition helps to decrease the required magnitude of the $M$ transition from $\Delta M\simeq -0.2$ to $\Delta M \simeq -0.1$. In the absence of the $w$ transition the Hubble tension could still be resolved via an $M$ transition with $\Delta M\simeq -0.2$ and no deformation of $D_L(z)$. Even though this approach would be simpler it would require a larger amplitude of the $\Delta M$ transition at $z_t\simeq 0.01$ while it would not address the abrupt feature in the Pantheon data shown in Fig. \ref{fig6}. Nevertheless, the simplicity of such an approach in an attractive feature and thus this model deserves a detailed investigation by comparing its predictions with current and future data.

\textbf{Numerical Analysis Files}: The numerical files for the reproduction of the figures can be found \href{https://github.com/GeorgeAlestas/LwMPT}{here}.

\section*{Acknowledgements}
We thank Savvas Nesseris and Valerio Marra for useful discussions. LK's research is co-financed by Greece and the European Union (European Social Fund- ESF) through the Operational Programme ``Human Resources Development, Education and Lifelong Learning" in the context of the project ``Strengthening Human Resources Research Potential via Doctorate Research – 2nd Cycle" (MIS-5000432), implemented by the State Scholarships Foundation (IKY). LP's and GA's research is co-financed by Greece and the European Union (European Social Fund - ESF) through the Operational Programme "Human Resources Development, Education and Lifelong Learning 2014-2020" in the context of the project  "Scalar fields in Curved Spacetimes: Soliton Solutions, Observational Results and Gravitational Waves" (MIS 5047648).

\appendix
\section{Data Used in the Analysis}
\label{sec:Appendix_A}
The covariance matrix which corresponds to the latest \plcdm CMB distance prior data (shift parameter $R$ and the acoustic scale $l_{a}$), for a flat universe has the following form \cite{Zhai:2018vmm}
\be
    C_{ij}=10^{-8}\times \begin{pmatrix}
     1598.9554~~  17112.007 \\
     17112.007~~  811208.45 \\
    \end{pmatrix} \nonumber
\ee
where the corresponding \plcdm values for $R$ and $l_a$ are presented in Table \ref{tab:CMBdat}. Furthermore, we present the full dataset of the BAO and CC likelihoods used in the Mathematica analysis in Tables \ref{tab:data-bao} and \ref{tab:data-cc} respectively.

\begin{widetext}

\begin{longtable}{|c|c|c|c|}
\caption{The CMB Distance Prior data for a flat Universe used in our analysis.}
\label{tab:CMBdat}\\
\hline
Index & CMB Observable & CMB Value & Reference\\
\hline
1 & $R$   & $1.74963 $ & \cite{Zhai:2018vmm}\\ 
2 & $l_a$ & $301.80845$ & \cite{Zhai:2018vmm}\\
\hline
\end{longtable}

\begin{longtable}{ | c | c | c | c | c | c |  }
\caption{The BAO data that have been used in the analysis along with the corresponding references.}
\label{tab:data-bao}\\
% header and footer information
\hline
   Index & $z$ & $D_A /r_s$ (Mpc) & $D_{H} /r_s (km/sec \cdot Mpc)$ & $D_V /r_s$ (Mpc) & Ref.  \\
\hline
1 & $0.106$ & - & - & $2.98 \pm 0.13 $ & \cite{Beutler:2011hx} \\
2 & $0.44$ & - & - & $13.69 \pm 5.82$ & \cite{Blake:2012pj} \\
3 & $0.6$ & - & - & $13.77 \pm 3.11$ & \cite{Blake:2012pj} \\
4 & $0.73$ & -& - & $16.89 \pm 5.28$ & \cite{Blake:2012pj} \\
5 & $2.34$ & $11.28 \pm 0.65$ & - & - & \cite{Agathe:2019vsu} \\
6 & $2.34$ & - & $9.18 \pm 0.28$ & -  & \cite{Agathe:2019vsu} \\
7 & $0.15$ & - & - & $4.465 \pm 0.168$  & \cite{Ross:2014qpa} \\
8 & $0.32$ & - & - & $8.62 \pm 0.15$  & \cite{Anderson:2013zyy} \\
9 & $0.57$ & - & - & $13.7 \pm 0.12$  & \cite{Anderson:2013zyy} \\
\hline
\end{longtable}

\begin{longtable}{|c|c|c|c|}
\caption{The Cosmic Chronometer data that have been used in the analysis.}
\label{tab:data-cc}\\
% header and footer information
\hline
   Index & $z$ & $H(z) (km/sec \cdot  Mpc)$ & Ref.  \\
\hline
1 & $0.09$ & $69 \pm 12$ & \cite{Jimenez2003Feb} \\
2 & $0.17$ & $83 \pm 8$ & \cite{PhysRevD.71.123001} \\
3 & $0.179$ & $75 \pm 4$  & \cite{Moresco_2012} \\
4 & $0.199$ & $75 \pm 5$  & \cite{Moresco_2012} \\
5 & $0.27$ & $77 \pm 14$  & \cite{PhysRevD.71.123001} \\
6 & $0.352$ & $83 \pm 14$  & \cite{Moresco_2012} \\
7 & $0.3802$ & $83 \pm 13.5$  & \cite{Moresco2016May} \\
8 & $0.4$ & $95 \pm 17$  & \cite{PhysRevD.71.123001} \\
9 & $0.4004$ & $77 \pm 10.2$  & \cite{Moresco2016May} \\
10 & $0.4247$ & $87.1 \pm 11.2$  & \cite{Moresco2016May} \\
11 & $0.4497$ & $92.8 \pm 12.9$  & \cite{Moresco2016May} \\
12 & $0.4783$ & $80.9 \pm 9$  & \cite{Moresco2016May} \\
13 & $0.48$ & $97 \pm 62$  & \cite{Stern2010Feb} \\
14 & $0.593$ & $104 \pm 13$  & \cite{Moresco_2012} \\
15 & $0.68$ & $92 \pm 8$ & \cite{Moresco_2012} \\
16 & $0.781$ & $105 \pm 12$  & \cite{Moresco_2012} \\
17 & $0.875$ & $125 \pm 17$  & \cite{Zhang2012Jul} \\
18 & $0.88$ & $90 \pm 40$  & \cite{Stern2010Feb} \\
19 & $0.9$ & $117 \pm 23$  & \cite{PhysRevD.71.123001} \\
20 & $1.037$ & $154 \pm 20$  & \cite{Moresco_2012} \\
21 & $1.3$ & $168 \pm 17$  & \cite{PhysRevD.71.123001} \\
22 & $1.363$ & $160 \pm 33.6$  & \cite{Moresco2015Jun} \\
23 & $1.43$ & $177 \pm 18$  & \cite{PhysRevD.71.123001} \\
24 & $1.53$ & $140 \pm 14$  & \cite{PhysRevD.71.123001} \\
25 & $1.75$ & $202 \pm 40$  & \cite{PhysRevD.71.123001} \\
26 & $1.965$ & $186.5 \pm 50.4$  & \cite{Moresco2015Jun} \\
27 & $0.35$ & $82.7 \pm 8.4$  & \cite{Chuang2013Oct} \\
28 & $0.44$ & $82.6 \pm 7.8$  & \cite{Blake2012Sep} \\
29 & $0.57$ & $96.8 \pm 3.4$  & \cite{Anderson:2013zyy} \\
30 & $0.6$ & $87.9 \pm 6.1$  & \cite{Blake2012Sep} \\
31 & $0.73$ & $97.3 \pm 7$  & \cite{Blake2012Sep} \\
32 & $2.34$ & $222 \pm 7$  & \cite{Delubac2015Feb} \\
33 & $0.07$ & $69 \pm 19.6$  & \cite{Zhang2012Jul} \\
34 & $0.12$ & $68.6 \pm 26.2$  & \cite{Zhang2012Jul} \\
35 & $0.2$ & $72.9 \pm 29.6$  & \cite{Zhang2012Jul} \\
36 & $0.24$ & $79.69 \pm 2.65$  & \cite{Gazta_aga_2009} \\
37 & $0.28$ & $88.8 \pm 36.6$  & \cite{Zhang2012Jul} \\
38 & $0.43$ & $86.45 \pm 3.68$  & \cite{Gazta_aga_2009} \\
39 & $0.57$ & $92.4 \pm 4.5$  & \cite{Samushia2013Feb} \\
40 & $2.3$ & $224 \pm 8$  & \cite{Busca2013Apr} \\
41 & $2.36$ & $226 \pm 8$  & \cite{Font_Ribera_2014} \\
\hline
\end{longtable}
\end{widetext}

\raggedleft
\bibliography{Bibliography}

%merlin.mbs apsrev4-1.bst 2010-07-25 4.21a (PWD, AO, DPC) hacked
%Control: key (0)
%Control: author (0) dotless jnrlst
%Control: editor formatted (1) identically to author
%Control: production of article title (0) allowed
%Control: page (1) range
%Control: year (0) verbatim
%Control: production of eprint (0) enabled
\begin{thebibliography}{96}%
\makeatletter
\providecommand \@ifxundefined [1]{%
 \@ifx{#1\undefined}
}%
\providecommand \@ifnum [1]{%
 \ifnum #1\expandafter \@firstoftwo
 \else \expandafter \@secondoftwo
 \fi
}%
\providecommand \@ifx [1]{%
 \ifx #1\expandafter \@firstoftwo
 \else \expandafter \@secondoftwo
 \fi
}%
\providecommand \natexlab [1]{#1}%
\providecommand \enquote  [1]{``#1''}%
\providecommand \bibnamefont  [1]{#1}%
\providecommand \bibfnamefont [1]{#1}%
\providecommand \citenamefont [1]{#1}%
\providecommand \href@noop [0]{\@secondoftwo}%
\providecommand \href [0]{\begingroup \@sanitize@url \@href}%
\providecommand \@href[1]{\@@startlink{#1}\@@href}%
\providecommand \@@href[1]{\endgroup#1\@@endlink}%
\providecommand \@sanitize@url [0]{\catcode `\\12\catcode `\$12\catcode
  `\&12\catcode `\#12\catcode `\^12\catcode `\_12\catcode `\%12\relax}%
\providecommand \@@startlink[1]{}%
\providecommand \@@endlink[0]{}%
\providecommand \url  [0]{\begingroup\@sanitize@url \@url }%
\providecommand \@url [1]{\endgroup\@href {#1}{\urlprefix }}%
\providecommand \urlprefix  [0]{URL }%
\providecommand \Eprint [0]{\href }%
\providecommand \doibase [0]{http://dx.doi.org/}%
\providecommand \selectlanguage [0]{\@gobble}%
\providecommand \bibinfo  [0]{\@secondoftwo}%
\providecommand \bibfield  [0]{\@secondoftwo}%
\providecommand \translation [1]{[#1]}%
\providecommand \BibitemOpen [0]{}%
\providecommand \bibitemStop [0]{}%
\providecommand \bibitemNoStop [0]{.\EOS\space}%
\providecommand \EOS [0]{\spacefactor3000\relax}%
\providecommand \BibitemShut  [1]{\csname bibitem#1\endcsname}%
\let\auto@bib@innerbib\@empty
%</preamble>
\bibitem [{\citenamefont {Riess}\ \emph {et~al.}(2019)\citenamefont {Riess},
  \citenamefont {Casertano}, \citenamefont {Yuan}, \citenamefont {Macri},\ and\
  \citenamefont {Scolnic}}]{Riess:2019cxk}%
  \BibitemOpen
  \bibfield  {author} {\bibinfo {author} {\bibfnamefont {Adam~G.}\ \bibnamefont
  {Riess}}, \bibinfo {author} {\bibfnamefont {Stefano}\ \bibnamefont
  {Casertano}}, \bibinfo {author} {\bibfnamefont {Wenlong}\ \bibnamefont
  {Yuan}}, \bibinfo {author} {\bibfnamefont {Lucas~M.}\ \bibnamefont {Macri}},
  \ and\ \bibinfo {author} {\bibfnamefont {Dan}\ \bibnamefont {Scolnic}},\
  }\bibfield  {title} {\enquote {\bibinfo {title} {{Large Magellanic Cloud
  Cepheid Standards Provide a 1\% Foundation for the Determination of the
  Hubble Constant and Stronger Evidence for Physics beyond $\Lambda$CDM}},}\
  }\href {\doibase 10.3847/1538-4357/ab1422} {\bibfield  {journal} {\bibinfo
  {journal} {Astrophys. J.}\ }\textbf {\bibinfo {volume} {876}},\ \bibinfo
  {pages} {85} (\bibinfo {year} {2019})},\ \Eprint
  {http://arxiv.org/abs/1903.07603} {arXiv:1903.07603 [astro-ph.CO]}
  \BibitemShut {NoStop}%
\bibitem [{\citenamefont {Riess}\ \emph {et~al.}(2018)\citenamefont {Riess}
  \emph {et~al.}}]{Riess:2018byc}%
  \BibitemOpen
  \bibfield  {author} {\bibinfo {author} {\bibfnamefont {Adam~G.}\ \bibnamefont
  {Riess}} \emph {et~al.},\ }\bibfield  {title} {\enquote {\bibinfo {title}
  {{Milky Way Cepheid Standards for Measuring Cosmic Distances and Application
  to Gaia DR2: Implications for the Hubble Constant}},}\ }\href {\doibase
  10.3847/1538-4357/aac82e} {\bibfield  {journal} {\bibinfo  {journal}
  {Astrophys. J.}\ }\textbf {\bibinfo {volume} {861}},\ \bibinfo {pages} {126}
  (\bibinfo {year} {2018})},\ \Eprint {http://arxiv.org/abs/1804.10655}
  {arXiv:1804.10655 [astro-ph.CO]} \BibitemShut {NoStop}%
\bibitem [{\citenamefont {Riess}\ \emph {et~al.}(2016)\citenamefont {Riess}
  \emph {et~al.}}]{Riess:2016jrr}%
  \BibitemOpen
  \bibfield  {author} {\bibinfo {author} {\bibfnamefont {Adam~G.}\ \bibnamefont
  {Riess}} \emph {et~al.},\ }\bibfield  {title} {\enquote {\bibinfo {title} {{A
  2.4\% Determination of the Local Value of the Hubble Constant}},}\ }\href
  {\doibase 10.3847/0004-637X/826/1/56} {\bibfield  {journal} {\bibinfo
  {journal} {Astrophys. J.}\ }\textbf {\bibinfo {volume} {826}},\ \bibinfo
  {pages} {56} (\bibinfo {year} {2016})},\ \Eprint
  {http://arxiv.org/abs/1604.01424} {arXiv:1604.01424 [astro-ph.CO]}
  \BibitemShut {NoStop}%
\bibitem [{\citenamefont {Freedman}\ \emph {et~al.}(2020)\citenamefont
  {Freedman}, \citenamefont {Madore}, \citenamefont {Hoyt}, \citenamefont
  {Jang}, \citenamefont {Beaton}, \citenamefont {Lee}, \citenamefont {Monson},
  \citenamefont {Neeley},\ and\ \citenamefont {Rich}}]{Freedman:2020dne}%
  \BibitemOpen
  \bibfield  {author} {\bibinfo {author} {\bibfnamefont {Wendy~L.}\
  \bibnamefont {Freedman}}, \bibinfo {author} {\bibfnamefont {Barry~F.}\
  \bibnamefont {Madore}}, \bibinfo {author} {\bibfnamefont {Taylor}\
  \bibnamefont {Hoyt}}, \bibinfo {author} {\bibfnamefont {In~Sung}\
  \bibnamefont {Jang}}, \bibinfo {author} {\bibfnamefont {Rachael}\
  \bibnamefont {Beaton}}, \bibinfo {author} {\bibfnamefont {Myung~Gyoon}\
  \bibnamefont {Lee}}, \bibinfo {author} {\bibfnamefont {Andrew}\ \bibnamefont
  {Monson}}, \bibinfo {author} {\bibfnamefont {Jill}\ \bibnamefont {Neeley}}, \
  and\ \bibinfo {author} {\bibfnamefont {Jeffrey}\ \bibnamefont {Rich}},\
  }\bibfield  {title} {\enquote {\bibinfo {title} {{Calibration of the Tip of
  the Red Giant Branch (TRGB)}},}\ }\href {\doibase 10.3847/1538-4357/ab7339}
  {\  (\bibinfo {year} {2020}),\ 10.3847/1538-4357/ab7339},\ \Eprint
  {http://arxiv.org/abs/2002.01550} {arXiv:2002.01550 [astro-ph.GA]}
  \BibitemShut {NoStop}%
\bibitem [{\citenamefont {Freedman}\ \emph {et~al.}(2019)\citenamefont
  {Freedman} \emph {et~al.}}]{Freedman:2019jwv}%
  \BibitemOpen
  \bibfield  {author} {\bibinfo {author} {\bibfnamefont {Wendy~L.}\
  \bibnamefont {Freedman}} \emph {et~al.},\ }\bibfield  {title} {\enquote
  {\bibinfo {title} {{The Carnegie-Chicago Hubble Program. VIII. An Independent
  Determination of the Hubble Constant Based on the Tip of the Red Giant
  Branch}},}\ }\href {\doibase 10.3847/1538-4357/ab2f73} {\  (\bibinfo {year}
  {2019}),\ 10.3847/1538-4357/ab2f73},\ \Eprint
  {http://arxiv.org/abs/1907.05922} {arXiv:1907.05922 [astro-ph.CO]}
  \BibitemShut {NoStop}%
\bibitem [{\citenamefont {Pesce}\ \emph {et~al.}(2020)\citenamefont {Pesce}
  \emph {et~al.}}]{Pesce:2020xfe}%
  \BibitemOpen
  \bibfield  {author} {\bibinfo {author} {\bibfnamefont {D.W.}\ \bibnamefont
  {Pesce}} \emph {et~al.},\ }\bibfield  {title} {\enquote {\bibinfo {title}
  {{The Megamaser Cosmology Project. XIII. Combined Hubble constant
  constraints}},}\ }\href {\doibase 10.3847/2041-8213/ab75f0} {\bibfield
  {journal} {\bibinfo  {journal} {Astrophys. J. Lett.}\ }\textbf {\bibinfo
  {volume} {891}},\ \bibinfo {pages} {L1} (\bibinfo {year} {2020})},\ \Eprint
  {http://arxiv.org/abs/2001.09213} {arXiv:2001.09213 [astro-ph.CO]}
  \BibitemShut {NoStop}%
\bibitem [{\citenamefont {Aghanim}\ \emph {et~al.}(2020)\citenamefont {Aghanim}
  \emph {et~al.}}]{Aghanim:2018eyx}%
  \BibitemOpen
  \bibfield  {author} {\bibinfo {author} {\bibfnamefont {N.}~\bibnamefont
  {Aghanim}} \emph {et~al.} (\bibinfo {collaboration} {Planck}),\ }\bibfield
  {title} {\enquote {\bibinfo {title} {{Planck 2018 results. VI. Cosmological
  parameters}},}\ }\href {\doibase 10.1051/0004-6361/201833910} {\bibfield
  {journal} {\bibinfo  {journal} {Astron. Astrophys.}\ }\textbf {\bibinfo
  {volume} {641}},\ \bibinfo {pages} {A6} (\bibinfo {year} {2020})},\ \Eprint
  {http://arxiv.org/abs/1807.06209} {arXiv:1807.06209 [astro-ph.CO]}
  \BibitemShut {NoStop}%
\bibitem [{\citenamefont {Addison}\ \emph {et~al.}(2018)\citenamefont
  {Addison}, \citenamefont {Watts}, \citenamefont {Bennett}, \citenamefont
  {Halpern}, \citenamefont {Hinshaw},\ and\ \citenamefont
  {Weiland}}]{Addison:2017fdm}%
  \BibitemOpen
  \bibfield  {author} {\bibinfo {author} {\bibfnamefont {G.E.}\ \bibnamefont
  {Addison}}, \bibinfo {author} {\bibfnamefont {D.J.}\ \bibnamefont {Watts}},
  \bibinfo {author} {\bibfnamefont {C.L.}\ \bibnamefont {Bennett}}, \bibinfo
  {author} {\bibfnamefont {M.}~\bibnamefont {Halpern}}, \bibinfo {author}
  {\bibfnamefont {G.}~\bibnamefont {Hinshaw}}, \ and\ \bibinfo {author}
  {\bibfnamefont {J.L.}\ \bibnamefont {Weiland}},\ }\bibfield  {title}
  {\enquote {\bibinfo {title} {{Elucidating $\Lambda$CDM: Impact of Baryon
  Acoustic Oscillation Measurements on the Hubble Constant Discrepancy}},}\
  }\href {\doibase 10.3847/1538-4357/aaa1ed} {\bibfield  {journal} {\bibinfo
  {journal} {Astrophys. J.}\ }\textbf {\bibinfo {volume} {853}},\ \bibinfo
  {pages} {119} (\bibinfo {year} {2018})},\ \Eprint
  {http://arxiv.org/abs/1707.06547} {arXiv:1707.06547 [astro-ph.CO]}
  \BibitemShut {NoStop}%
\bibitem [{\citenamefont {Sch\"oneberg}\ \emph {et~al.}(2019)\citenamefont
  {Sch\"oneberg}, \citenamefont {Lesgourgues},\ and\ \citenamefont
  {Hooper}}]{Schoneberg:2019wmt}%
  \BibitemOpen
  \bibfield  {author} {\bibinfo {author} {\bibfnamefont {Nils}\ \bibnamefont
  {Sch\"oneberg}}, \bibinfo {author} {\bibfnamefont {Julien}\ \bibnamefont
  {Lesgourgues}}, \ and\ \bibinfo {author} {\bibfnamefont {Deanna~C.}\
  \bibnamefont {Hooper}},\ }\bibfield  {title} {\enquote {\bibinfo {title}
  {{The BAO+BBN take on the Hubble tension}},}\ }\href {\doibase
  10.1088/1475-7516/2019/10/029} {\bibfield  {journal} {\bibinfo  {journal}
  {JCAP}\ }\textbf {\bibinfo {volume} {10}},\ \bibinfo {pages} {029} (\bibinfo
  {year} {2019})},\ \Eprint {http://arxiv.org/abs/1907.11594} {arXiv:1907.11594
  [astro-ph.CO]} \BibitemShut {NoStop}%
\bibitem [{\citenamefont {Birrer}\ \emph {et~al.}(2019)\citenamefont {Birrer}
  \emph {et~al.}}]{Birrer:2018vtm}%
  \BibitemOpen
  \bibfield  {author} {\bibinfo {author} {\bibfnamefont {S.}~\bibnamefont
  {Birrer}} \emph {et~al.},\ }\bibfield  {title} {\enquote {\bibinfo {title}
  {{H0LiCOW - IX. Cosmographic analysis of the doubly imaged quasar SDSS
  1206+4332 and a new measurement of the Hubble constant}},}\ }\href {\doibase
  10.1093/mnras/stz200} {\bibfield  {journal} {\bibinfo  {journal} {Mon. Not.
  Roy. Astron. Soc.}\ }\textbf {\bibinfo {volume} {484}},\ \bibinfo {pages}
  {4726} (\bibinfo {year} {2019})},\ \Eprint {http://arxiv.org/abs/1809.01274}
  {arXiv:1809.01274 [astro-ph.CO]} \BibitemShut {NoStop}%
\bibitem [{\citenamefont {Shajib}\ \emph {et~al.}(2020)\citenamefont {Shajib}
  \emph {et~al.}}]{Shajib:2019toy}%
  \BibitemOpen
  \bibfield  {author} {\bibinfo {author} {\bibfnamefont {A.J.}\ \bibnamefont
  {Shajib}} \emph {et~al.} (\bibinfo {collaboration} {DES}),\ }\bibfield
  {title} {\enquote {\bibinfo {title} {{STRIDES: a 3.9 per cent measurement of
  the Hubble constant from the strong lens system DES
  J0408\ensuremath{-}5354}},}\ }\href {\doibase 10.1093/mnras/staa828}
  {\bibfield  {journal} {\bibinfo  {journal} {Mon. Not. Roy. Astron. Soc.}\
  }\textbf {\bibinfo {volume} {494}},\ \bibinfo {pages} {6072--6102} (\bibinfo
  {year} {2020})},\ \Eprint {http://arxiv.org/abs/1910.06306} {arXiv:1910.06306
  [astro-ph.CO]} \BibitemShut {NoStop}%
\bibitem [{\citenamefont {Abbott}\ \emph {et~al.}(2019)\citenamefont {Abbott}
  \emph {et~al.}}]{Abbott:2019yzh}%
  \BibitemOpen
  \bibfield  {author} {\bibinfo {author} {\bibfnamefont {B.P.}\ \bibnamefont
  {Abbott}} \emph {et~al.} (\bibinfo {collaboration} {LIGO Scientific,
  Virgo}),\ }\bibfield  {title} {\enquote {\bibinfo {title} {{A
  gravitational-wave measurement of the Hubble constant following the second
  observing run of Advanced LIGO and Virgo}},}\ }\href@noop {} {\  (\bibinfo
  {year} {2019})},\ \Eprint {http://arxiv.org/abs/1908.06060} {arXiv:1908.06060
  [astro-ph.CO]} \BibitemShut {NoStop}%
\bibitem [{\citenamefont {Soares-Santos}\ \emph {et~al.}(2019)\citenamefont
  {Soares-Santos} \emph {et~al.}}]{Soares-Santos:2019irc}%
  \BibitemOpen
  \bibfield  {author} {\bibinfo {author} {\bibfnamefont {M.}~\bibnamefont
  {Soares-Santos}} \emph {et~al.} (\bibinfo {collaboration} {DES, LIGO
  Scientific, Virgo}),\ }\bibfield  {title} {\enquote {\bibinfo {title} {{First
  Measurement of the Hubble Constant from a Dark Standard Siren using the Dark
  Energy Survey Galaxies and the LIGO/Virgo Binary\textendash{}Black-hole
  Merger GW170814}},}\ }\href {\doibase 10.3847/2041-8213/ab14f1} {\bibfield
  {journal} {\bibinfo  {journal} {Astrophys. J. Lett.}\ }\textbf {\bibinfo
  {volume} {876}},\ \bibinfo {pages} {L7} (\bibinfo {year} {2019})},\ \Eprint
  {http://arxiv.org/abs/1901.01540} {arXiv:1901.01540 [astro-ph.CO]}
  \BibitemShut {NoStop}%
\bibitem [{\citenamefont {Knox}\ and\ \citenamefont
  {Millea}(2020)}]{Knox:2019rjx}%
  \BibitemOpen
  \bibfield  {author} {\bibinfo {author} {\bibfnamefont {Lloyd}\ \bibnamefont
  {Knox}}\ and\ \bibinfo {author} {\bibfnamefont {Marius}\ \bibnamefont
  {Millea}},\ }\bibfield  {title} {\enquote {\bibinfo {title} {{Hubble constant
  hunter\textquoteright{}s guide}},}\ }\href {\doibase
  10.1103/PhysRevD.101.043533} {\bibfield  {journal} {\bibinfo  {journal}
  {Phys. Rev. D}\ }\textbf {\bibinfo {volume} {101}},\ \bibinfo {pages}
  {043533} (\bibinfo {year} {2020})},\ \Eprint
  {http://arxiv.org/abs/1908.03663} {arXiv:1908.03663 [astro-ph.CO]}
  \BibitemShut {NoStop}%
\bibitem [{\citenamefont {M\"ortsell}\ and\ \citenamefont
  {Dhawan}(2018)}]{Mortsell:2018mfj}%
  \BibitemOpen
  \bibfield  {author} {\bibinfo {author} {\bibfnamefont {Edvard}\ \bibnamefont
  {M\"ortsell}}\ and\ \bibinfo {author} {\bibfnamefont {Suhail}\ \bibnamefont
  {Dhawan}},\ }\bibfield  {title} {\enquote {\bibinfo {title} {{Does the Hubble
  constant tension call for new physics?}}}\ }\href {\doibase
  10.1088/1475-7516/2018/09/025} {\bibfield  {journal} {\bibinfo  {journal}
  {JCAP}\ }\textbf {\bibinfo {volume} {09}},\ \bibinfo {pages} {025} (\bibinfo
  {year} {2018})},\ \Eprint {http://arxiv.org/abs/1801.07260} {arXiv:1801.07260
  [astro-ph.CO]} \BibitemShut {NoStop}%
\bibitem [{\citenamefont {Efstathiou}(2020)}]{Efstathiou:2020wxn}%
  \BibitemOpen
  \bibfield  {author} {\bibinfo {author} {\bibfnamefont {G.}~\bibnamefont
  {Efstathiou}},\ }\bibfield  {title} {\enquote {\bibinfo {title} {{A Lockdown
  Perspective on the Hubble Tension (with comments from the SH0ES team)}},}\
  }\href@noop {} {\  (\bibinfo {year} {2020})},\ \Eprint
  {http://arxiv.org/abs/2007.10716} {arXiv:2007.10716 [astro-ph.CO]}
  \BibitemShut {NoStop}%
\bibitem [{\citenamefont {Kazantzidis}\ and\ \citenamefont
  {Perivolaropoulos}(2020)}]{Kazantzidis:2020tko}%
  \BibitemOpen
  \bibfield  {author} {\bibinfo {author} {\bibfnamefont {L.}~\bibnamefont
  {Kazantzidis}}\ and\ \bibinfo {author} {\bibfnamefont {L.}~\bibnamefont
  {Perivolaropoulos}},\ }\bibfield  {title} {\enquote {\bibinfo {title} {{Hints
  of a Local Matter Underdensity or Modified Gravity in the Low $z$ Pantheon
  data}},}\ }\href {\doibase 10.1103/PhysRevD.102.023520} {\bibfield  {journal}
  {\bibinfo  {journal} {Phys. Rev. D}\ }\textbf {\bibinfo {volume} {102}},\
  \bibinfo {pages} {023520} (\bibinfo {year} {2020})},\ \Eprint
  {http://arxiv.org/abs/2004.02155} {arXiv:2004.02155 [astro-ph.CO]}
  \BibitemShut {NoStop}%
\bibitem [{\citenamefont {Kazantzidis}\ \emph {et~al.}(2020)\citenamefont
  {Kazantzidis}, \citenamefont {Koo}, \citenamefont {Nesseris}, \citenamefont
  {Perivolaropoulos},\ and\ \citenamefont {Shafieloo}}]{Kazantzidis:2020xta}%
  \BibitemOpen
  \bibfield  {author} {\bibinfo {author} {\bibfnamefont {L.}~\bibnamefont
  {Kazantzidis}}, \bibinfo {author} {\bibfnamefont {H.}~\bibnamefont {Koo}},
  \bibinfo {author} {\bibfnamefont {S.}~\bibnamefont {Nesseris}}, \bibinfo
  {author} {\bibfnamefont {L.}~\bibnamefont {Perivolaropoulos}}, \ and\
  \bibinfo {author} {\bibfnamefont {A.}~\bibnamefont {Shafieloo}},\ }\bibfield
  {title} {\enquote {\bibinfo {title} {{Hints for possible low redshift
  oscillation around the best fit $\Lambda$CDM model in the expansion history
  of the universe}},}\ }\href {\doibase 10.1093/mnras/staa3866} {\  (\bibinfo
  {year} {2020}),\ 10.1093/mnras/staa3866},\ \Eprint
  {http://arxiv.org/abs/2010.03491} {arXiv:2010.03491 [astro-ph.CO]}
  \BibitemShut {NoStop}%
\bibitem [{\citenamefont {Sapone}\ \emph {et~al.}(2020)\citenamefont {Sapone},
  \citenamefont {Nesseris},\ and\ \citenamefont {Bengaly}}]{Sapone:2020wwz}%
  \BibitemOpen
  \bibfield  {author} {\bibinfo {author} {\bibfnamefont {Domenico}\
  \bibnamefont {Sapone}}, \bibinfo {author} {\bibfnamefont {Savvas}\
  \bibnamefont {Nesseris}}, \ and\ \bibinfo {author} {\bibfnamefont
  {Carlos~A.P.}\ \bibnamefont {Bengaly}},\ }\bibfield  {title} {\enquote
  {\bibinfo {title} {{Is there any measurable redshift dependence on the SN Ia
  absolute magnitude?}}}\ }\href@noop {} {\  (\bibinfo {year} {2020})},\
  \Eprint {http://arxiv.org/abs/2006.05461} {arXiv:2006.05461 [astro-ph.CO]}
  \BibitemShut {NoStop}%
\bibitem [{\citenamefont {Soltis}\ \emph {et~al.}(2020)\citenamefont {Soltis},
  \citenamefont {Casertano},\ and\ \citenamefont {Riess}}]{Soltis:2020gpl}%
  \BibitemOpen
  \bibfield  {author} {\bibinfo {author} {\bibfnamefont {John}\ \bibnamefont
  {Soltis}}, \bibinfo {author} {\bibfnamefont {Stefano}\ \bibnamefont
  {Casertano}}, \ and\ \bibinfo {author} {\bibfnamefont {Adam~G.}\ \bibnamefont
  {Riess}},\ }\bibfield  {title} {\enquote {\bibinfo {title} {{The Parallax of
  Omega Centauri Measured from Gaia EDR3 and a Direct, Geometric Calibration of
  the Tip of the Red Giant Branch and the Hubble Constant}},}\ }\href@noop {}
  {\  (\bibinfo {year} {2020})},\ \Eprint {http://arxiv.org/abs/2012.09196}
  {arXiv:2012.09196 [astro-ph.GA]} \BibitemShut {NoStop}%
\bibitem [{\citenamefont {Riess}\ \emph {et~al.}(2020)\citenamefont {Riess},
  \citenamefont {Casertano}, \citenamefont {Yuan}, \citenamefont {Bowers},
  \citenamefont {Macri}, \citenamefont {Zinn},\ and\ \citenamefont
  {Scolnic}}]{Riess:2020fzl}%
  \BibitemOpen
  \bibfield  {author} {\bibinfo {author} {\bibfnamefont {Adam~G.}\ \bibnamefont
  {Riess}}, \bibinfo {author} {\bibfnamefont {Stefano}\ \bibnamefont
  {Casertano}}, \bibinfo {author} {\bibfnamefont {Wenlong}\ \bibnamefont
  {Yuan}}, \bibinfo {author} {\bibfnamefont {J.~Bradley}\ \bibnamefont
  {Bowers}}, \bibinfo {author} {\bibfnamefont {Lucas}\ \bibnamefont {Macri}},
  \bibinfo {author} {\bibfnamefont {Joel~C.}\ \bibnamefont {Zinn}}, \ and\
  \bibinfo {author} {\bibfnamefont {Dan}\ \bibnamefont {Scolnic}},\ }\bibfield
  {title} {\enquote {\bibinfo {title} {{Cosmic Distances Calibrated to 1\%
  Precision with Gaia EDR3 Parallaxes and Hubble Space Telescope Photometry of
  75 Milky Way Cepheids Confirm Tension with LambdaCDM}},}\ }\href@noop {} {\
  (\bibinfo {year} {2020})},\ \Eprint {http://arxiv.org/abs/2012.08534}
  {arXiv:2012.08534 [astro-ph.CO]} \BibitemShut {NoStop}%
\bibitem [{\citenamefont {Blinov}\ \emph {et~al.}(2019)\citenamefont {Blinov},
  \citenamefont {Kelly}, \citenamefont {Krnjaic},\ and\ \citenamefont
  {McDermott}}]{Blinov:2019gcj}%
  \BibitemOpen
  \bibfield  {author} {\bibinfo {author} {\bibfnamefont {Nikita}\ \bibnamefont
  {Blinov}}, \bibinfo {author} {\bibfnamefont {Kevin~James}\ \bibnamefont
  {Kelly}}, \bibinfo {author} {\bibfnamefont {Gordan~Z}\ \bibnamefont
  {Krnjaic}}, \ and\ \bibinfo {author} {\bibfnamefont {Samuel~D}\ \bibnamefont
  {McDermott}},\ }\bibfield  {title} {\enquote {\bibinfo {title} {{Constraining
  the Self-Interacting Neutrino Interpretation of the Hubble Tension}},}\
  }\href {\doibase 10.1103/PhysRevLett.123.191102} {\bibfield  {journal}
  {\bibinfo  {journal} {Phys. Rev. Lett.}\ }\textbf {\bibinfo {volume} {123}},\
  \bibinfo {pages} {191102} (\bibinfo {year} {2019})},\ \Eprint
  {http://arxiv.org/abs/1905.02727} {arXiv:1905.02727 [astro-ph.CO]}
  \BibitemShut {NoStop}%
\bibitem [{\citenamefont {Poulin}\ \emph {et~al.}(2019)\citenamefont {Poulin},
  \citenamefont {Smith}, \citenamefont {Karwal},\ and\ \citenamefont
  {Kamionkowski}}]{Poulin:2018cxd}%
  \BibitemOpen
  \bibfield  {author} {\bibinfo {author} {\bibfnamefont {Vivian}\ \bibnamefont
  {Poulin}}, \bibinfo {author} {\bibfnamefont {Tristan~L.}\ \bibnamefont
  {Smith}}, \bibinfo {author} {\bibfnamefont {Tanvi}\ \bibnamefont {Karwal}}, \
  and\ \bibinfo {author} {\bibfnamefont {Marc}\ \bibnamefont {Kamionkowski}},\
  }\bibfield  {title} {\enquote {\bibinfo {title} {{Early Dark Energy Can
  Resolve The Hubble Tension}},}\ }\href {\doibase
  10.1103/PhysRevLett.122.221301} {\bibfield  {journal} {\bibinfo  {journal}
  {Phys. Rev. Lett.}\ }\textbf {\bibinfo {volume} {122}},\ \bibinfo {pages}
  {221301} (\bibinfo {year} {2019})},\ \Eprint
  {http://arxiv.org/abs/1811.04083} {arXiv:1811.04083 [astro-ph.CO]}
  \BibitemShut {NoStop}%
\bibitem [{\citenamefont {Sakstein}\ and\ \citenamefont
  {Trodden}(2020)}]{Sakstein:2019fmf}%
  \BibitemOpen
  \bibfield  {author} {\bibinfo {author} {\bibfnamefont {Jeremy}\ \bibnamefont
  {Sakstein}}\ and\ \bibinfo {author} {\bibfnamefont {Mark}\ \bibnamefont
  {Trodden}},\ }\bibfield  {title} {\enquote {\bibinfo {title} {{Early Dark
  Energy from Massive Neutrinos as a Natural Resolution of the Hubble
  Tension}},}\ }\href {\doibase 10.1103/PhysRevLett.124.161301} {\bibfield
  {journal} {\bibinfo  {journal} {Phys. Rev. Lett.}\ }\textbf {\bibinfo
  {volume} {124}},\ \bibinfo {pages} {161301} (\bibinfo {year} {2020})},\
  \Eprint {http://arxiv.org/abs/1911.11760} {arXiv:1911.11760 [astro-ph.CO]}
  \BibitemShut {NoStop}%
\bibitem [{\citenamefont {Agrawal}\ \emph {et~al.}(2019)\citenamefont
  {Agrawal}, \citenamefont {Cyr-Racine}, \citenamefont {Pinner},\ and\
  \citenamefont {Randall}}]{Agrawal:2019lmo}%
  \BibitemOpen
  \bibfield  {author} {\bibinfo {author} {\bibfnamefont {Prateek}\ \bibnamefont
  {Agrawal}}, \bibinfo {author} {\bibfnamefont {Francis-Yan}\ \bibnamefont
  {Cyr-Racine}}, \bibinfo {author} {\bibfnamefont {David}\ \bibnamefont
  {Pinner}}, \ and\ \bibinfo {author} {\bibfnamefont {Lisa}\ \bibnamefont
  {Randall}},\ }\bibfield  {title} {\enquote {\bibinfo {title} {{Rock 'n' Roll
  Solutions to the Hubble Tension}},}\ }\href@noop {} {\  (\bibinfo {year}
  {2019})},\ \Eprint {http://arxiv.org/abs/1904.01016} {arXiv:1904.01016
  [astro-ph.CO]} \BibitemShut {NoStop}%
\bibitem [{\citenamefont {Lin}\ \emph {et~al.}(2019)\citenamefont {Lin},
  \citenamefont {Benevento}, \citenamefont {Hu},\ and\ \citenamefont
  {Raveri}}]{Lin:2019qug}%
  \BibitemOpen
  \bibfield  {author} {\bibinfo {author} {\bibfnamefont {Meng-Xiang}\
  \bibnamefont {Lin}}, \bibinfo {author} {\bibfnamefont {Giampaolo}\
  \bibnamefont {Benevento}}, \bibinfo {author} {\bibfnamefont {Wayne}\
  \bibnamefont {Hu}}, \ and\ \bibinfo {author} {\bibfnamefont {Marco}\
  \bibnamefont {Raveri}},\ }\bibfield  {title} {\enquote {\bibinfo {title}
  {{Acoustic Dark Energy: Potential Conversion of the Hubble Tension}},}\
  }\href {\doibase 10.1103/PhysRevD.100.063542} {\bibfield  {journal} {\bibinfo
   {journal} {Phys. Rev. D}\ }\textbf {\bibinfo {volume} {100}},\ \bibinfo
  {pages} {063542} (\bibinfo {year} {2019})},\ \Eprint
  {http://arxiv.org/abs/1905.12618} {arXiv:1905.12618 [astro-ph.CO]}
  \BibitemShut {NoStop}%
\bibitem [{\citenamefont {Braglia}\ \emph {et~al.}(2020)\citenamefont
  {Braglia}, \citenamefont {Emond}, \citenamefont {Finelli}, \citenamefont
  {Gumrukcuoglu},\ and\ \citenamefont {Koyama}}]{Braglia:2020bym}%
  \BibitemOpen
  \bibfield  {author} {\bibinfo {author} {\bibfnamefont {Matteo}\ \bibnamefont
  {Braglia}}, \bibinfo {author} {\bibfnamefont {William~T.}\ \bibnamefont
  {Emond}}, \bibinfo {author} {\bibfnamefont {Fabio}\ \bibnamefont {Finelli}},
  \bibinfo {author} {\bibfnamefont {A.~Emir}\ \bibnamefont {Gumrukcuoglu}}, \
  and\ \bibinfo {author} {\bibfnamefont {Kazuya}\ \bibnamefont {Koyama}},\
  }\bibfield  {title} {\enquote {\bibinfo {title} {{Unified framework for Early
  Dark Energy from $\alpha$-attractors}},}\ }\href@noop {} {\  (\bibinfo {year}
  {2020})},\ \Eprint {http://arxiv.org/abs/2005.14053} {arXiv:2005.14053
  [astro-ph.CO]} \BibitemShut {NoStop}%
\bibitem [{\citenamefont {Niedermann}\ and\ \citenamefont
  {Sloth}(2020)}]{Niedermann:2020dwg}%
  \BibitemOpen
  \bibfield  {author} {\bibinfo {author} {\bibfnamefont {Florian}\ \bibnamefont
  {Niedermann}}\ and\ \bibinfo {author} {\bibfnamefont {Martin~S.}\
  \bibnamefont {Sloth}},\ }\bibfield  {title} {\enquote {\bibinfo {title}
  {{Resolving the Hubble Tension with New Early Dark Energy}},}\ }\href
  {\doibase 10.1103/PhysRevD.102.063527} {\bibfield  {journal} {\bibinfo
  {journal} {Phys. Rev. D}\ }\textbf {\bibinfo {volume} {102}},\ \bibinfo
  {pages} {063527} (\bibinfo {year} {2020})},\ \Eprint
  {http://arxiv.org/abs/2006.06686} {arXiv:2006.06686 [astro-ph.CO]}
  \BibitemShut {NoStop}%
\bibitem [{\citenamefont {Smith}\ \emph
  {et~al.}(2020{\natexlab{a}})\citenamefont {Smith}, \citenamefont {Poulin},
  \citenamefont {Bernal}, \citenamefont {Boddy}, \citenamefont {Kamionkowski},\
  and\ \citenamefont {Murgia}}]{Smith:2020rxx}%
  \BibitemOpen
  \bibfield  {author} {\bibinfo {author} {\bibfnamefont {Tristan~L.}\
  \bibnamefont {Smith}}, \bibinfo {author} {\bibfnamefont {Vivian}\
  \bibnamefont {Poulin}}, \bibinfo {author} {\bibfnamefont {Jos\'e~Luis}\
  \bibnamefont {Bernal}}, \bibinfo {author} {\bibfnamefont {Kimberly~K.}\
  \bibnamefont {Boddy}}, \bibinfo {author} {\bibfnamefont {Marc}\ \bibnamefont
  {Kamionkowski}}, \ and\ \bibinfo {author} {\bibfnamefont {Riccardo}\
  \bibnamefont {Murgia}},\ }\bibfield  {title} {\enquote {\bibinfo {title}
  {{Early dark energy is not excluded by current large-scale structure
  data}},}\ }\href@noop {} {\  (\bibinfo {year} {2020}{\natexlab{a}})},\
  \Eprint {http://arxiv.org/abs/2009.10740} {arXiv:2009.10740 [astro-ph.CO]}
  \BibitemShut {NoStop}%
\bibitem [{\citenamefont {Benisty}(2019)}]{Benisty:2019pxb}%
  \BibitemOpen
  \bibfield  {author} {\bibinfo {author} {\bibfnamefont {David}\ \bibnamefont
  {Benisty}},\ }\bibfield  {title} {\enquote {\bibinfo {title} {{Decaying
  coupled Fermions to curvature and the $H_0$ tension}},}\ }\href@noop {} {\
  (\bibinfo {year} {2019})},\ \Eprint {http://arxiv.org/abs/1912.11124}
  {arXiv:1912.11124 [gr-qc]} \BibitemShut {NoStop}%
\bibitem [{\citenamefont {Ballardini}\ \emph {et~al.}(2020)\citenamefont
  {Ballardini}, \citenamefont {Braglia}, \citenamefont {Finelli}, \citenamefont
  {Paoletti}, \citenamefont {Starobinsky},\ and\ \citenamefont
  {Umilt\`a}}]{Ballardini:2020iws}%
  \BibitemOpen
  \bibfield  {author} {\bibinfo {author} {\bibfnamefont {Mario}\ \bibnamefont
  {Ballardini}}, \bibinfo {author} {\bibfnamefont {Matteo}\ \bibnamefont
  {Braglia}}, \bibinfo {author} {\bibfnamefont {Fabio}\ \bibnamefont
  {Finelli}}, \bibinfo {author} {\bibfnamefont {Daniela}\ \bibnamefont
  {Paoletti}}, \bibinfo {author} {\bibfnamefont {Alexei~A.}\ \bibnamefont
  {Starobinsky}}, \ and\ \bibinfo {author} {\bibfnamefont {Caterina}\
  \bibnamefont {Umilt\`a}},\ }\bibfield  {title} {\enquote {\bibinfo {title}
  {{Scalar-tensor theories of gravity, neutrino physics, and the $H_0$
  tension}},}\ }\href {\doibase 10.1088/1475-7516/2020/10/044} {\bibfield
  {journal} {\bibinfo  {journal} {JCAP}\ }\textbf {\bibinfo {volume} {10}},\
  \bibinfo {pages} {044} (\bibinfo {year} {2020})},\ \Eprint
  {http://arxiv.org/abs/2004.14349} {arXiv:2004.14349 [astro-ph.CO]}
  \BibitemShut {NoStop}%
\bibitem [{\citenamefont {D'Amico}\ \emph {et~al.}(2020)\citenamefont
  {D'Amico}, \citenamefont {Senatore}, \citenamefont {Zhang},\ and\
  \citenamefont {Zheng}}]{DAmico:2020ods}%
  \BibitemOpen
  \bibfield  {author} {\bibinfo {author} {\bibfnamefont {Guido}\ \bibnamefont
  {D'Amico}}, \bibinfo {author} {\bibfnamefont {Leonardo}\ \bibnamefont
  {Senatore}}, \bibinfo {author} {\bibfnamefont {Pierre}\ \bibnamefont
  {Zhang}}, \ and\ \bibinfo {author} {\bibfnamefont {Henry}\ \bibnamefont
  {Zheng}},\ }\bibfield  {title} {\enquote {\bibinfo {title} {{The Hubble
  Tension in Light of the Full-Shape Analysis of Large-Scale Structure
  Data}},}\ }\href@noop {} {\  (\bibinfo {year} {2020})},\ \Eprint
  {http://arxiv.org/abs/2006.12420} {arXiv:2006.12420 [astro-ph.CO]}
  \BibitemShut {NoStop}%
\bibitem [{\citenamefont {Haridasu}\ \emph {et~al.}(2020)\citenamefont
  {Haridasu}, \citenamefont {Viel},\ and\ \citenamefont
  {Vittorio}}]{Haridasu:2020pms}%
  \BibitemOpen
  \bibfield  {author} {\bibinfo {author} {\bibfnamefont {Balakrishna~S.}\
  \bibnamefont {Haridasu}}, \bibinfo {author} {\bibfnamefont {Matteo}\
  \bibnamefont {Viel}}, \ and\ \bibinfo {author} {\bibfnamefont {Nicola}\
  \bibnamefont {Vittorio}},\ }\bibfield  {title} {\enquote {\bibinfo {title}
  {{Sources of $H_0$-tensions in dark energy scenarios}},}\ }\href@noop {} {\
  (\bibinfo {year} {2020})},\ \Eprint {http://arxiv.org/abs/2012.10324}
  {arXiv:2012.10324 [astro-ph.CO]} \BibitemShut {NoStop}%
\bibitem [{\citenamefont {Krishnan}\ \emph
  {et~al.}(2020{\natexlab{a}})\citenamefont {Krishnan}, \citenamefont
  {Colg\'ain}, \citenamefont {Ruchika}, \citenamefont {Sen}, \citenamefont
  {Sheikh-Jabbari},\ and\ \citenamefont {Yang}}]{Krishnan:2020obg}%
  \BibitemOpen
  \bibfield  {author} {\bibinfo {author} {\bibfnamefont {C.}~\bibnamefont
  {Krishnan}}, \bibinfo {author} {\bibfnamefont {Eoin~\'O.}\ \bibnamefont
  {Colg\'ain}}, \bibinfo {author} {\bibnamefont {Ruchika}}, \bibinfo {author}
  {\bibfnamefont {Anjan~A.}\ \bibnamefont {Sen}}, \bibinfo {author}
  {\bibfnamefont {M.M.}\ \bibnamefont {Sheikh-Jabbari}}, \ and\ \bibinfo
  {author} {\bibfnamefont {Tao}\ \bibnamefont {Yang}},\ }\bibfield  {title}
  {\enquote {\bibinfo {title} {{Is there an early Universe solution to Hubble
  tension?}}}\ }\href {\doibase 10.1103/PhysRevD.102.103525} {\bibfield
  {journal} {\bibinfo  {journal} {Phys. Rev. D}\ }\textbf {\bibinfo {volume}
  {102}},\ \bibinfo {pages} {103525} (\bibinfo {year} {2020}{\natexlab{a}})},\
  \Eprint {http://arxiv.org/abs/2002.06044} {arXiv:2002.06044 [astro-ph.CO]}
  \BibitemShut {NoStop}%
\bibitem [{\citenamefont {Hildebrandt}\ \emph {et~al.}(2017)\citenamefont
  {Hildebrandt} \emph {et~al.}}]{Hildebrandt:2016iqg}%
  \BibitemOpen
  \bibfield  {author} {\bibinfo {author} {\bibfnamefont {H.}~\bibnamefont
  {Hildebrandt}} \emph {et~al.},\ }\bibfield  {title} {\enquote {\bibinfo
  {title} {{KiDS-450: Cosmological parameter constraints from tomographic weak
  gravitational lensing}},}\ }\href {\doibase 10.1093/mnras/stw2805} {\bibfield
   {journal} {\bibinfo  {journal} {Mon. Not. Roy. Astron. Soc.}\ }\textbf
  {\bibinfo {volume} {465}},\ \bibinfo {pages} {1454} (\bibinfo {year}
  {2017})},\ \Eprint {http://arxiv.org/abs/1606.05338} {arXiv:1606.05338
  [astro-ph.CO]} \BibitemShut {NoStop}%
\bibitem [{\citenamefont {Joudaki}\ \emph {et~al.}(2018)\citenamefont {Joudaki}
  \emph {et~al.}}]{Joudaki:2017zdt}%
  \BibitemOpen
  \bibfield  {author} {\bibinfo {author} {\bibfnamefont {Shahab}\ \bibnamefont
  {Joudaki}} \emph {et~al.},\ }\bibfield  {title} {\enquote {\bibinfo {title}
  {{KiDS-450 + 2dFLenS: Cosmological parameter constraints from weak
  gravitational lensing tomography and overlapping redshift-space galaxy
  clustering}},}\ }\href {\doibase 10.1093/mnras/stx2820} {\bibfield  {journal}
  {\bibinfo  {journal} {Mon. Not. Roy. Astron. Soc.}\ }\textbf {\bibinfo
  {volume} {474}},\ \bibinfo {pages} {4894--4924} (\bibinfo {year} {2018})},\
  \Eprint {http://arxiv.org/abs/1707.06627} {arXiv:1707.06627 [astro-ph.CO]}
  \BibitemShut {NoStop}%
\bibitem [{\citenamefont {K\"ohlinger}\ \emph {et~al.}(2017)\citenamefont
  {K\"ohlinger} \emph {et~al.}}]{Kohlinger:2017sxk}%
  \BibitemOpen
  \bibfield  {author} {\bibinfo {author} {\bibfnamefont {F.}~\bibnamefont
  {K\"ohlinger}} \emph {et~al.},\ }\bibfield  {title} {\enquote {\bibinfo
  {title} {{KiDS-450: The tomographic weak lensing power spectrum and
  constraints on cosmological parameters}},}\ }\href {\doibase
  10.1093/mnras/stx1820} {\bibfield  {journal} {\bibinfo  {journal} {Mon. Not.
  Roy. Astron. Soc.}\ }\textbf {\bibinfo {volume} {471}},\ \bibinfo {pages}
  {4412--4435} (\bibinfo {year} {2017})},\ \Eprint
  {http://arxiv.org/abs/1706.02892} {arXiv:1706.02892 [astro-ph.CO]}
  \BibitemShut {NoStop}%
\bibitem [{\citenamefont {Abbott}\ \emph {et~al.}(2018)\citenamefont {Abbott}
  \emph {et~al.}}]{Abbott:2017wau}%
  \BibitemOpen
  \bibfield  {author} {\bibinfo {author} {\bibfnamefont {T.M.C.}\ \bibnamefont
  {Abbott}} \emph {et~al.} (\bibinfo {collaboration} {DES}),\ }\bibfield
  {title} {\enquote {\bibinfo {title} {{Dark Energy Survey year 1 results:
  Cosmological constraints from galaxy clustering and weak lensing}},}\ }\href
  {\doibase 10.1103/PhysRevD.98.043526} {\bibfield  {journal} {\bibinfo
  {journal} {Phys. Rev. D}\ }\textbf {\bibinfo {volume} {98}},\ \bibinfo
  {pages} {043526} (\bibinfo {year} {2018})},\ \Eprint
  {http://arxiv.org/abs/1708.01530} {arXiv:1708.01530 [astro-ph.CO]}
  \BibitemShut {NoStop}%
\bibitem [{\citenamefont {Macaulay}\ \emph {et~al.}(2013)\citenamefont
  {Macaulay}, \citenamefont {Wehus},\ and\ \citenamefont
  {Eriksen}}]{Macaulay:2013swa}%
  \BibitemOpen
  \bibfield  {author} {\bibinfo {author} {\bibfnamefont {Edward}\ \bibnamefont
  {Macaulay}}, \bibinfo {author} {\bibfnamefont {Ingunn~Kathrine}\ \bibnamefont
  {Wehus}}, \ and\ \bibinfo {author} {\bibfnamefont {Hans~Kristian}\
  \bibnamefont {Eriksen}},\ }\bibfield  {title} {\enquote {\bibinfo {title}
  {{Lower Growth Rate from Recent Redshift Space Distortion Measurements than
  Expected from Planck}},}\ }\href {\doibase 10.1103/PhysRevLett.111.161301}
  {\bibfield  {journal} {\bibinfo  {journal} {Phys. Rev. Lett.}\ }\textbf
  {\bibinfo {volume} {111}},\ \bibinfo {pages} {161301} (\bibinfo {year}
  {2013})},\ \Eprint {http://arxiv.org/abs/1303.6583} {arXiv:1303.6583
  [astro-ph.CO]} \BibitemShut {NoStop}%
\bibitem [{\citenamefont {Kazantzidis}\ and\ \citenamefont
  {Perivolaropoulos}(2018)}]{Kazantzidis:2018rnb}%
  \BibitemOpen
  \bibfield  {author} {\bibinfo {author} {\bibfnamefont {Lavrentios}\
  \bibnamefont {Kazantzidis}}\ and\ \bibinfo {author} {\bibfnamefont
  {Leandros}\ \bibnamefont {Perivolaropoulos}},\ }\bibfield  {title} {\enquote
  {\bibinfo {title} {{Evolution of the $f\sigma_8$ tension with the
  Planck15/$\Lambda$CDM determination and implications for modified gravity
  theories}},}\ }\href {\doibase 10.1103/PhysRevD.97.103503} {\bibfield
  {journal} {\bibinfo  {journal} {Phys. Rev. D}\ }\textbf {\bibinfo {volume}
  {97}},\ \bibinfo {pages} {103503} (\bibinfo {year} {2018})},\ \Eprint
  {http://arxiv.org/abs/1803.01337} {arXiv:1803.01337 [astro-ph.CO]}
  \BibitemShut {NoStop}%
\bibitem [{\citenamefont {Nesseris}\ \emph {et~al.}(2017)\citenamefont
  {Nesseris}, \citenamefont {Pantazis},\ and\ \citenamefont
  {Perivolaropoulos}}]{Nesseris:2017vor}%
  \BibitemOpen
  \bibfield  {author} {\bibinfo {author} {\bibfnamefont {Savvas}\ \bibnamefont
  {Nesseris}}, \bibinfo {author} {\bibfnamefont {George}\ \bibnamefont
  {Pantazis}}, \ and\ \bibinfo {author} {\bibfnamefont {Leandros}\ \bibnamefont
  {Perivolaropoulos}},\ }\bibfield  {title} {\enquote {\bibinfo {title}
  {{Tension and constraints on modified gravity parametrizations of
  $G_{\textrm{eff}}(z)$ from growth rate and Planck data}},}\ }\href {\doibase
  10.1103/PhysRevD.96.023542} {\bibfield  {journal} {\bibinfo  {journal} {Phys.
  Rev. D}\ }\textbf {\bibinfo {volume} {96}},\ \bibinfo {pages} {023542}
  (\bibinfo {year} {2017})},\ \Eprint {http://arxiv.org/abs/1703.10538}
  {arXiv:1703.10538 [astro-ph.CO]} \BibitemShut {NoStop}%
\bibitem [{\citenamefont {Skara}\ and\ \citenamefont
  {Perivolaropoulos}(2020)}]{Skara:2019usd}%
  \BibitemOpen
  \bibfield  {author} {\bibinfo {author} {\bibfnamefont {F.}~\bibnamefont
  {Skara}}\ and\ \bibinfo {author} {\bibfnamefont {L.}~\bibnamefont
  {Perivolaropoulos}},\ }\bibfield  {title} {\enquote {\bibinfo {title}
  {{Tension of the $E_G$ statistic and redshift space distortion data with the
  Planck - $\Lambda CDM$ model and implications for weakening gravity}},}\
  }\href {\doibase 10.1103/PhysRevD.101.063521} {\bibfield  {journal} {\bibinfo
   {journal} {Phys. Rev. D}\ }\textbf {\bibinfo {volume} {101}},\ \bibinfo
  {pages} {063521} (\bibinfo {year} {2020})},\ \Eprint
  {http://arxiv.org/abs/1911.10609} {arXiv:1911.10609 [astro-ph.CO]}
  \BibitemShut {NoStop}%
\bibitem [{\citenamefont {Kazantzidis}\ and\ \citenamefont
  {Perivolaropoulos}(2019)}]{Kazantzidis:2019dvk}%
  \BibitemOpen
  \bibfield  {author} {\bibinfo {author} {\bibfnamefont {Lavrentios}\
  \bibnamefont {Kazantzidis}}\ and\ \bibinfo {author} {\bibfnamefont
  {Leandros}\ \bibnamefont {Perivolaropoulos}},\ }\bibfield  {title} {\enquote
  {\bibinfo {title} {{Is gravity getting weaker at low z? Observational
  evidence and theoretical implications}},}\ }\href@noop {} {\  (\bibinfo
  {year} {2019})},\ \Eprint {http://arxiv.org/abs/1907.03176} {arXiv:1907.03176
  [astro-ph.CO]} \BibitemShut {NoStop}%
\bibitem [{\citenamefont {Kazantzidis}\ \emph {et~al.}(2019)\citenamefont
  {Kazantzidis}, \citenamefont {Perivolaropoulos},\ and\ \citenamefont
  {Skara}}]{Kazantzidis:2018jtb}%
  \BibitemOpen
  \bibfield  {author} {\bibinfo {author} {\bibfnamefont {L.}~\bibnamefont
  {Kazantzidis}}, \bibinfo {author} {\bibfnamefont {L.}~\bibnamefont
  {Perivolaropoulos}}, \ and\ \bibinfo {author} {\bibfnamefont
  {F.}~\bibnamefont {Skara}},\ }\bibfield  {title} {\enquote {\bibinfo {title}
  {{Constraining power of cosmological observables: blind redshift spots and
  optimal ranges}},}\ }\href {\doibase 10.1103/PhysRevD.99.063537} {\bibfield
  {journal} {\bibinfo  {journal} {Phys. Rev. D}\ }\textbf {\bibinfo {volume}
  {99}},\ \bibinfo {pages} {063537} (\bibinfo {year} {2019})},\ \Eprint
  {http://arxiv.org/abs/1812.05356} {arXiv:1812.05356 [astro-ph.CO]}
  \BibitemShut {NoStop}%
\bibitem [{\citenamefont {Di~Valentino}\ \emph {et~al.}(2021)\citenamefont
  {Di~Valentino}, \citenamefont {Mena}, \citenamefont {Pan}, \citenamefont
  {Visinelli}, \citenamefont {Yang}, \citenamefont {Melchiorri}, \citenamefont
  {Mota}, \citenamefont {Riess},\ and\ \citenamefont
  {Silk}}]{DiValentino:2021izs}%
  \BibitemOpen
  \bibfield  {author} {\bibinfo {author} {\bibfnamefont {Eleonora}\
  \bibnamefont {Di~Valentino}}, \bibinfo {author} {\bibfnamefont {Olga}\
  \bibnamefont {Mena}}, \bibinfo {author} {\bibfnamefont {Supriya}\
  \bibnamefont {Pan}}, \bibinfo {author} {\bibfnamefont {Luca}\ \bibnamefont
  {Visinelli}}, \bibinfo {author} {\bibfnamefont {Weiqiang}\ \bibnamefont
  {Yang}}, \bibinfo {author} {\bibfnamefont {Alessandro}\ \bibnamefont
  {Melchiorri}}, \bibinfo {author} {\bibfnamefont {David~F.}\ \bibnamefont
  {Mota}}, \bibinfo {author} {\bibfnamefont {Adam~G.}\ \bibnamefont {Riess}}, \
  and\ \bibinfo {author} {\bibfnamefont {Joseph}\ \bibnamefont {Silk}},\
  }\bibfield  {title} {\enquote {\bibinfo {title} {{In the Realm of the Hubble
  tension $-$ a Review of Solutions}},}\ }\href@noop {} {\  (\bibinfo {year}
  {2021})},\ \Eprint {http://arxiv.org/abs/2103.01183} {arXiv:2103.01183
  [astro-ph.CO]} \BibitemShut {NoStop}%
\bibitem [{\citenamefont {Alestas}\ \emph {et~al.}(2020)\citenamefont
  {Alestas}, \citenamefont {Kazantzidis},\ and\ \citenamefont
  {Perivolaropoulos}}]{Alestas:2020mvb}%
  \BibitemOpen
  \bibfield  {author} {\bibinfo {author} {\bibfnamefont {G.}~\bibnamefont
  {Alestas}}, \bibinfo {author} {\bibfnamefont {L.}~\bibnamefont
  {Kazantzidis}}, \ and\ \bibinfo {author} {\bibfnamefont {L.}~\bibnamefont
  {Perivolaropoulos}},\ }\bibfield  {title} {\enquote {\bibinfo {title} {{$H_0$
  tension, phantom dark energy, and cosmological parameter degeneracies}},}\
  }\href {\doibase 10.1103/PhysRevD.101.123516} {\bibfield  {journal} {\bibinfo
   {journal} {Phys. Rev. D}\ }\textbf {\bibinfo {volume} {101}},\ \bibinfo
  {pages} {123516} (\bibinfo {year} {2020})},\ \Eprint
  {http://arxiv.org/abs/2004.08363} {arXiv:2004.08363 [astro-ph.CO]}
  \BibitemShut {NoStop}%
\bibitem [{\citenamefont {Di~Valentino}\ \emph {et~al.}(2016)\citenamefont
  {Di~Valentino}, \citenamefont {Melchiorri},\ and\ \citenamefont
  {Silk}}]{DiValentino:2016hlg}%
  \BibitemOpen
  \bibfield  {author} {\bibinfo {author} {\bibfnamefont {Eleonora}\
  \bibnamefont {Di~Valentino}}, \bibinfo {author} {\bibfnamefont {Alessandro}\
  \bibnamefont {Melchiorri}}, \ and\ \bibinfo {author} {\bibfnamefont {Joseph}\
  \bibnamefont {Silk}},\ }\bibfield  {title} {\enquote {\bibinfo {title}
  {{Reconciling Planck with the local value of $H_0$ in extended parameter
  space}},}\ }\href {\doibase 10.1016/j.physletb.2016.08.043} {\bibfield
  {journal} {\bibinfo  {journal} {Phys. Lett. B}\ }\textbf {\bibinfo {volume}
  {761}},\ \bibinfo {pages} {242--246} (\bibinfo {year} {2016})},\ \Eprint
  {http://arxiv.org/abs/1606.00634} {arXiv:1606.00634 [astro-ph.CO]}
  \BibitemShut {NoStop}%
\bibitem [{\citenamefont {Smith}\ \emph
  {et~al.}(2020{\natexlab{b}})\citenamefont {Smith}, \citenamefont {Poulin},\
  and\ \citenamefont {Amin}}]{Smith:2019ihp}%
  \BibitemOpen
  \bibfield  {author} {\bibinfo {author} {\bibfnamefont {Tristan~L.}\
  \bibnamefont {Smith}}, \bibinfo {author} {\bibfnamefont {Vivian}\
  \bibnamefont {Poulin}}, \ and\ \bibinfo {author} {\bibfnamefont {Mustafa~A.}\
  \bibnamefont {Amin}},\ }\bibfield  {title} {\enquote {\bibinfo {title}
  {{Oscillating scalar fields and the Hubble tension: a resolution with novel
  signatures}},}\ }\href {\doibase 10.1103/PhysRevD.101.063523} {\bibfield
  {journal} {\bibinfo  {journal} {Phys. Rev. D}\ }\textbf {\bibinfo {volume}
  {101}},\ \bibinfo {pages} {063523} (\bibinfo {year} {2020}{\natexlab{b}})},\
  \Eprint {http://arxiv.org/abs/1908.06995} {arXiv:1908.06995 [astro-ph.CO]}
  \BibitemShut {NoStop}%
\bibitem [{\citenamefont {Vagnozzi}(2020)}]{Vagnozzi:2019ezj}%
  \BibitemOpen
  \bibfield  {author} {\bibinfo {author} {\bibfnamefont {Sunny}\ \bibnamefont
  {Vagnozzi}},\ }\bibfield  {title} {\enquote {\bibinfo {title} {{New physics
  in light of the $H_0$ tension: An alternative view}},}\ }\href {\doibase
  10.1103/PhysRevD.102.023518} {\bibfield  {journal} {\bibinfo  {journal}
  {Phys. Rev. D}\ }\textbf {\bibinfo {volume} {102}},\ \bibinfo {pages}
  {023518} (\bibinfo {year} {2020})},\ \Eprint
  {http://arxiv.org/abs/1907.07569} {arXiv:1907.07569 [astro-ph.CO]}
  \BibitemShut {NoStop}%
\bibitem [{\citenamefont {Li}\ and\ \citenamefont
  {Shafieloo}(2019)}]{Li:2019yem}%
  \BibitemOpen
  \bibfield  {author} {\bibinfo {author} {\bibfnamefont {Xiaolei}\ \bibnamefont
  {Li}}\ and\ \bibinfo {author} {\bibfnamefont {Arman}\ \bibnamefont
  {Shafieloo}},\ }\bibfield  {title} {\enquote {\bibinfo {title} {{A Simple
  Phenomenological Emergent Dark Energy Model can Resolve the Hubble
  Tension}},}\ }\href {\doibase 10.3847/2041-8213/ab3e09} {\bibfield  {journal}
  {\bibinfo  {journal} {Astrophys. J. Lett.}\ }\textbf {\bibinfo {volume}
  {883}},\ \bibinfo {pages} {L3} (\bibinfo {year} {2019})},\ \Eprint
  {http://arxiv.org/abs/1906.08275} {arXiv:1906.08275 [astro-ph.CO]}
  \BibitemShut {NoStop}%
\bibitem [{\citenamefont {Di~Valentino}\ \emph {et~al.}(2020)\citenamefont
  {Di~Valentino}, \citenamefont {Mukherjee},\ and\ \citenamefont
  {Sen}}]{DiValentino:2020naf}%
  \BibitemOpen
  \bibfield  {author} {\bibinfo {author} {\bibfnamefont {Eleonora}\
  \bibnamefont {Di~Valentino}}, \bibinfo {author} {\bibfnamefont {Ankan}\
  \bibnamefont {Mukherjee}}, \ and\ \bibinfo {author} {\bibfnamefont
  {Anjan~A.}\ \bibnamefont {Sen}},\ }\bibfield  {title} {\enquote {\bibinfo
  {title} {{Dark Energy with Phantom Crossing and the $H_0$ tension}},}\
  }\href@noop {} {\  (\bibinfo {year} {2020})},\ \Eprint
  {http://arxiv.org/abs/2005.12587} {arXiv:2005.12587 [astro-ph.CO]}
  \BibitemShut {NoStop}%
\bibitem [{\citenamefont {Krishnan}\ \emph
  {et~al.}(2020{\natexlab{b}})\citenamefont {Krishnan}, \citenamefont
  {Colgain}, \citenamefont {Sheikh-Jabbari},\ and\ \citenamefont
  {Yang}}]{Krishnan:2020vaf}%
  \BibitemOpen
  \bibfield  {author} {\bibinfo {author} {\bibfnamefont {Chethan}\ \bibnamefont
  {Krishnan}}, \bibinfo {author} {\bibfnamefont {Eoin~O.}\ \bibnamefont
  {Colgain}}, \bibinfo {author} {\bibfnamefont {M.M.}\ \bibnamefont
  {Sheikh-Jabbari}}, \ and\ \bibinfo {author} {\bibfnamefont {Tao}\
  \bibnamefont {Yang}},\ }\bibfield  {title} {\enquote {\bibinfo {title}
  {{Running Hubble Tension and a H0 Diagnostic}},}\ }\href@noop {} {\
  (\bibinfo {year} {2020}{\natexlab{b}})},\ \Eprint
  {http://arxiv.org/abs/2011.02858} {arXiv:2011.02858 [astro-ph.CO]}
  \BibitemShut {NoStop}%
\bibitem [{\citenamefont {Mortonson}\ \emph {et~al.}(2009)\citenamefont
  {Mortonson}, \citenamefont {Hu},\ and\ \citenamefont
  {Huterer}}]{Mortonson:2009qq}%
  \BibitemOpen
  \bibfield  {author} {\bibinfo {author} {\bibfnamefont {Michael~J.}\
  \bibnamefont {Mortonson}}, \bibinfo {author} {\bibfnamefont {Wayne}\
  \bibnamefont {Hu}}, \ and\ \bibinfo {author} {\bibfnamefont {Dragan}\
  \bibnamefont {Huterer}},\ }\bibfield  {title} {\enquote {\bibinfo {title}
  {{Hiding dark energy transitions at low redshift}},}\ }\href {\doibase
  10.1103/PhysRevD.80.067301} {\bibfield  {journal} {\bibinfo  {journal} {Phys.
  Rev. D}\ }\textbf {\bibinfo {volume} {80}},\ \bibinfo {pages} {067301}
  (\bibinfo {year} {2009})},\ \Eprint {http://arxiv.org/abs/0908.1408}
  {arXiv:0908.1408 [astro-ph.CO]} \BibitemShut {NoStop}%
\bibitem [{\citenamefont {Benevento}\ \emph {et~al.}(2020)\citenamefont
  {Benevento}, \citenamefont {Hu},\ and\ \citenamefont
  {Raveri}}]{Benevento:2020fev}%
  \BibitemOpen
  \bibfield  {author} {\bibinfo {author} {\bibfnamefont {Giampaolo}\
  \bibnamefont {Benevento}}, \bibinfo {author} {\bibfnamefont {Wayne}\
  \bibnamefont {Hu}}, \ and\ \bibinfo {author} {\bibfnamefont {Marco}\
  \bibnamefont {Raveri}},\ }\bibfield  {title} {\enquote {\bibinfo {title}
  {{Can Late Dark Energy Transitions Raise the Hubble constant?}}}\ }\href
  {\doibase 10.1103/PhysRevD.101.103517} {\bibfield  {journal} {\bibinfo
  {journal} {Phys. Rev. D}\ }\textbf {\bibinfo {volume} {101}},\ \bibinfo
  {pages} {103517} (\bibinfo {year} {2020})},\ \Eprint
  {http://arxiv.org/abs/2002.11707} {arXiv:2002.11707 [astro-ph.CO]}
  \BibitemShut {NoStop}%
\bibitem [{\citenamefont {Dhawan}\ \emph {et~al.}(2020)\citenamefont {Dhawan},
  \citenamefont {Brout}, \citenamefont {Scolnic}, \citenamefont {Goobar},
  \citenamefont {Riess},\ and\ \citenamefont {Miranda}}]{Dhawan:2020xmp}%
  \BibitemOpen
  \bibfield  {author} {\bibinfo {author} {\bibfnamefont {S.}~\bibnamefont
  {Dhawan}}, \bibinfo {author} {\bibfnamefont {D.}~\bibnamefont {Brout}},
  \bibinfo {author} {\bibfnamefont {D.}~\bibnamefont {Scolnic}}, \bibinfo
  {author} {\bibfnamefont {A.}~\bibnamefont {Goobar}}, \bibinfo {author}
  {\bibfnamefont {A.G.}\ \bibnamefont {Riess}}, \ and\ \bibinfo {author}
  {\bibfnamefont {V.}~\bibnamefont {Miranda}},\ }\bibfield  {title} {\enquote
  {\bibinfo {title} {{Cosmological Model Insensitivity of Local $H_0$ from the
  Cepheid Distance Ladder}},}\ }\href {\doibase 10.3847/1538-4357/ab7fb0}
  {\bibfield  {journal} {\bibinfo  {journal} {Astrophys. J.}\ }\textbf
  {\bibinfo {volume} {894}},\ \bibinfo {pages} {54} (\bibinfo {year} {2020})},\
  \Eprint {http://arxiv.org/abs/2001.09260} {arXiv:2001.09260 [astro-ph.CO]}
  \BibitemShut {NoStop}%
\bibitem [{\citenamefont {Keeley}\ \emph {et~al.}(2019)\citenamefont {Keeley},
  \citenamefont {Joudaki}, \citenamefont {Kaplinghat},\ and\ \citenamefont
  {Kirkby}}]{Keeley:2019esp}%
  \BibitemOpen
  \bibfield  {author} {\bibinfo {author} {\bibfnamefont {Ryan~E.}\ \bibnamefont
  {Keeley}}, \bibinfo {author} {\bibfnamefont {Shahab}\ \bibnamefont
  {Joudaki}}, \bibinfo {author} {\bibfnamefont {Manoj}\ \bibnamefont
  {Kaplinghat}}, \ and\ \bibinfo {author} {\bibfnamefont {David}\ \bibnamefont
  {Kirkby}},\ }\bibfield  {title} {\enquote {\bibinfo {title} {{Implications of
  a transition in the dark energy equation of state for the $H_0$ and
  $\sigma_8$ tensions}},}\ }\href {\doibase 10.1088/1475-7516/2019/12/035}
  {\bibfield  {journal} {\bibinfo  {journal} {JCAP}\ }\textbf {\bibinfo
  {volume} {12}},\ \bibinfo {pages} {035} (\bibinfo {year} {2019})},\ \Eprint
  {http://arxiv.org/abs/1905.10198} {arXiv:1905.10198 [astro-ph.CO]}
  \BibitemShut {NoStop}%
\bibitem [{\citenamefont {Bassett}\ \emph {et~al.}(2002)\citenamefont
  {Bassett}, \citenamefont {Kunz}, \citenamefont {Silk},\ and\ \citenamefont
  {Ungarelli}}]{Bassett:2002qu}%
  \BibitemOpen
  \bibfield  {author} {\bibinfo {author} {\bibfnamefont {Bruce~A.}\
  \bibnamefont {Bassett}}, \bibinfo {author} {\bibfnamefont {Martin}\
  \bibnamefont {Kunz}}, \bibinfo {author} {\bibfnamefont {Joseph}\ \bibnamefont
  {Silk}}, \ and\ \bibinfo {author} {\bibfnamefont {Carlo}\ \bibnamefont
  {Ungarelli}},\ }\bibfield  {title} {\enquote {\bibinfo {title} {{A Late time
  transition in the cosmic dark energy?}}}\ }\href {\doibase
  10.1046/j.1365-8711.2002.05887.x} {\bibfield  {journal} {\bibinfo  {journal}
  {Mon. Not. Roy. Astron. Soc.}\ }\textbf {\bibinfo {volume} {336}},\ \bibinfo
  {pages} {1217--1222} (\bibinfo {year} {2002})},\ \Eprint
  {http://arxiv.org/abs/astro-ph/0203383} {arXiv:astro-ph/0203383} \BibitemShut
  {NoStop}%
\bibitem [{\citenamefont {Camarena}\ and\ \citenamefont
  {Marra}(2021)}]{Camarena:2021jlr}%
  \BibitemOpen
  \bibfield  {author} {\bibinfo {author} {\bibfnamefont {David}\ \bibnamefont
  {Camarena}}\ and\ \bibinfo {author} {\bibfnamefont {Valerio}\ \bibnamefont
  {Marra}},\ }\bibfield  {title} {\enquote {\bibinfo {title} {{Hockey-stick
  dark energy is not a solution to the $H_0$ crisis}},}\ }\href@noop {} {\
  (\bibinfo {year} {2021})},\ \Eprint {http://arxiv.org/abs/2101.08641}
  {arXiv:2101.08641 [astro-ph.CO]} \BibitemShut {NoStop}%
\bibitem [{\citenamefont {Camarena}\ and\ \citenamefont
  {Marra}(2020)}]{Camarena:2019moy}%
  \BibitemOpen
  \bibfield  {author} {\bibinfo {author} {\bibfnamefont {David}\ \bibnamefont
  {Camarena}}\ and\ \bibinfo {author} {\bibfnamefont {Valerio}\ \bibnamefont
  {Marra}},\ }\bibfield  {title} {\enquote {\bibinfo {title} {{Local
  determination of the Hubble constant and the deceleration parameter}},}\
  }\href {\doibase 10.1103/PhysRevResearch.2.013028} {\bibfield  {journal}
  {\bibinfo  {journal} {Phys. Rev. Res.}\ }\textbf {\bibinfo {volume} {2}},\
  \bibinfo {pages} {013028} (\bibinfo {year} {2020})},\ \Eprint
  {http://arxiv.org/abs/1906.11814} {arXiv:1906.11814 [astro-ph.CO]}
  \BibitemShut {NoStop}%
\bibitem [{\citenamefont {Scolnic}\ \emph {et~al.}(2018)\citenamefont {Scolnic}
  \emph {et~al.}}]{Scolnic:2017caz}%
  \BibitemOpen
  \bibfield  {author} {\bibinfo {author} {\bibfnamefont {D.M.}\ \bibnamefont
  {Scolnic}} \emph {et~al.},\ }\bibfield  {title} {\enquote {\bibinfo {title}
  {{The Complete Light-curve Sample of Spectroscopically Confirmed SNe Ia from
  Pan-STARRS1 and Cosmological Constraints from the Combined Pantheon
  Sample}},}\ }\href {\doibase 10.3847/1538-4357/aab9bb} {\bibfield  {journal}
  {\bibinfo  {journal} {Astrophys. J.}\ }\textbf {\bibinfo {volume} {859}},\
  \bibinfo {pages} {101} (\bibinfo {year} {2018})},\ \Eprint
  {http://arxiv.org/abs/1710.00845} {arXiv:1710.00845 [astro-ph.CO]}
  \BibitemShut {NoStop}%
\bibitem [{\citenamefont {Elgaroy}\ and\ \citenamefont
  {Multamaki}(2007)}]{Elgaroy:2007bv}%
  \BibitemOpen
  \bibfield  {author} {\bibinfo {author} {\bibfnamefont {Oystein}\ \bibnamefont
  {Elgaroy}}\ and\ \bibinfo {author} {\bibfnamefont {Tuomas}\ \bibnamefont
  {Multamaki}},\ }\bibfield  {title} {\enquote {\bibinfo {title} {{On using the
  CMB shift parameter in tests of models of dark energy}},}\ }\href {\doibase
  10.1051/0004-6361:20077292} {\bibfield  {journal} {\bibinfo  {journal}
  {Astron. Astrophys.}\ }\textbf {\bibinfo {volume} {471}},\ \bibinfo {pages}
  {65} (\bibinfo {year} {2007})},\ \Eprint
  {http://arxiv.org/abs/astro-ph/0702343} {arXiv:astro-ph/0702343} \BibitemShut
  {NoStop}%
\bibitem [{\citenamefont {Zhai}\ and\ \citenamefont
  {Wang}(2019)}]{Zhai:2018vmm}%
  \BibitemOpen
  \bibfield  {author} {\bibinfo {author} {\bibfnamefont {Zhongxu}\ \bibnamefont
  {Zhai}}\ and\ \bibinfo {author} {\bibfnamefont {Yun}\ \bibnamefont {Wang}},\
  }\bibfield  {title} {\enquote {\bibinfo {title} {{Robust and
  model-independent cosmological constraints from distance measurements}},}\
  }\href {\doibase 10.1088/1475-7516/2019/07/005} {\bibfield  {journal}
  {\bibinfo  {journal} {JCAP}\ }\textbf {\bibinfo {volume} {07}},\ \bibinfo
  {pages} {005} (\bibinfo {year} {2019})},\ \Eprint
  {http://arxiv.org/abs/1811.07425} {arXiv:1811.07425 [astro-ph.CO]}
  \BibitemShut {NoStop}%
\bibitem [{\citenamefont {Arjona}\ \emph {et~al.}(2019)\citenamefont {Arjona},
  \citenamefont {Cardona},\ and\ \citenamefont {Nesseris}}]{Arjona:2018jhh}%
  \BibitemOpen
  \bibfield  {author} {\bibinfo {author} {\bibfnamefont {Rub\'en}\ \bibnamefont
  {Arjona}}, \bibinfo {author} {\bibfnamefont {Wilmar}\ \bibnamefont
  {Cardona}}, \ and\ \bibinfo {author} {\bibfnamefont {Savvas}\ \bibnamefont
  {Nesseris}},\ }\bibfield  {title} {\enquote {\bibinfo {title} {{Unraveling
  the effective fluid approach for $f(R)$ models in the subhorizon
  approximation}},}\ }\href {\doibase 10.1103/PhysRevD.99.043516} {\bibfield
  {journal} {\bibinfo  {journal} {Phys. Rev. D}\ }\textbf {\bibinfo {volume}
  {99}},\ \bibinfo {pages} {043516} (\bibinfo {year} {2019})},\ \Eprint
  {http://arxiv.org/abs/1811.02469} {arXiv:1811.02469 [astro-ph.CO]}
  \BibitemShut {NoStop}%
\bibitem [{\citenamefont {Alestas}\ and\ \citenamefont
  {Perivolaropoulos}(2021)}]{Alestas:2021xes}%
  \BibitemOpen
  \bibfield  {author} {\bibinfo {author} {\bibfnamefont {G.}~\bibnamefont
  {Alestas}}\ and\ \bibinfo {author} {\bibfnamefont {L.}~\bibnamefont
  {Perivolaropoulos}},\ }\bibfield  {title} {\enquote {\bibinfo {title} {{Late
  time approaches to the Hubble tension deforming $H(z)$, worsen the growth
  tension}},}\ }\href@noop {} {\  (\bibinfo {year} {2021})},\ \Eprint
  {http://arxiv.org/abs/2103.04045} {arXiv:2103.04045 [astro-ph.CO]}
  \BibitemShut {NoStop}%
\bibitem [{\citenamefont {Marra}\ and\ \citenamefont
  {Perivolaropoulos}(2021)}]{Marra:2021fvf}%
  \BibitemOpen
  \bibfield  {author} {\bibinfo {author} {\bibfnamefont {Valerio}\ \bibnamefont
  {Marra}}\ and\ \bibinfo {author} {\bibfnamefont {Leandros}\ \bibnamefont
  {Perivolaropoulos}},\ }\bibfield  {title} {\enquote {\bibinfo {title} {{A
  rapid transition of $G_{\rm eff}$ at $z_t \simeq 0.01$ as a solution of the
  Hubble and growth tensions}},}\ }\href@noop {} {\  (\bibinfo {year}
  {2021})},\ \Eprint {http://arxiv.org/abs/2102.06012} {arXiv:2102.06012
  [astro-ph.CO]} \BibitemShut {NoStop}%
\bibitem [{\citenamefont {Wright}\ and\ \citenamefont
  {Li}(2018)}]{Wright:2017rsu}%
  \BibitemOpen
  \bibfield  {author} {\bibinfo {author} {\bibfnamefont {Bill~S.}\ \bibnamefont
  {Wright}}\ and\ \bibinfo {author} {\bibfnamefont {Baojiu}\ \bibnamefont
  {Li}},\ }\bibfield  {title} {\enquote {\bibinfo {title} {{Type Ia supernovae,
  standardizable candles, and gravity}},}\ }\href {\doibase
  10.1103/PhysRevD.97.083505} {\bibfield  {journal} {\bibinfo  {journal} {Phys.
  Rev. D}\ }\textbf {\bibinfo {volume} {97}},\ \bibinfo {pages} {083505}
  (\bibinfo {year} {2018})},\ \Eprint {http://arxiv.org/abs/1710.07018}
  {arXiv:1710.07018 [astro-ph.CO]} \BibitemShut {NoStop}%
\bibitem [{\citenamefont {Amendola}\ \emph {et~al.}(1999)\citenamefont
  {Amendola}, \citenamefont {Corasaniti},\ and\ \citenamefont
  {Occhionero}}]{Amendola:1999vu}%
  \BibitemOpen
  \bibfield  {author} {\bibinfo {author} {\bibfnamefont {Luca}\ \bibnamefont
  {Amendola}}, \bibinfo {author} {\bibfnamefont {Pier~Stefano}\ \bibnamefont
  {Corasaniti}}, \ and\ \bibinfo {author} {\bibfnamefont {Franco}\ \bibnamefont
  {Occhionero}},\ }\bibfield  {title} {\enquote {\bibinfo {title} {{Time
  variability of the gravitational constant and type Ia supernovae}},}\
  }\href@noop {} {\  (\bibinfo {year} {1999})},\ \Eprint
  {http://arxiv.org/abs/astro-ph/9907222} {arXiv:astro-ph/9907222} \BibitemShut
  {NoStop}%
\bibitem [{\citenamefont {Gaztanaga}\ \emph {et~al.}(2002)\citenamefont
  {Gaztanaga}, \citenamefont {Garcia-Berro}, \citenamefont {Isern},
  \citenamefont {Bravo},\ and\ \citenamefont {Dominguez}}]{Gaztanaga:2001fh}%
  \BibitemOpen
  \bibfield  {author} {\bibinfo {author} {\bibfnamefont {E.}~\bibnamefont
  {Gaztanaga}}, \bibinfo {author} {\bibfnamefont {E.}~\bibnamefont
  {Garcia-Berro}}, \bibinfo {author} {\bibfnamefont {J.}~\bibnamefont {Isern}},
  \bibinfo {author} {\bibfnamefont {E.}~\bibnamefont {Bravo}}, \ and\ \bibinfo
  {author} {\bibfnamefont {I.}~\bibnamefont {Dominguez}},\ }\bibfield  {title}
  {\enquote {\bibinfo {title} {{Bounds on the possible evolution of the
  gravitational constant from cosmological type Ia supernovae}},}\ }\href
  {\doibase 10.1103/PhysRevD.65.023506} {\bibfield  {journal} {\bibinfo
  {journal} {Phys. Rev. D}\ }\textbf {\bibinfo {volume} {65}},\ \bibinfo
  {pages} {023506} (\bibinfo {year} {2002})},\ \Eprint
  {http://arxiv.org/abs/astro-ph/0109299} {arXiv:astro-ph/0109299} \BibitemShut
  {NoStop}%
\bibitem [{\citenamefont {Gannouji}\ \emph {et~al.}(2018)\citenamefont
  {Gannouji}, \citenamefont {Kazantzidis}, \citenamefont {Perivolaropoulos},\
  and\ \citenamefont {Polarski}}]{Gannouji:2018ncm}%
  \BibitemOpen
  \bibfield  {author} {\bibinfo {author} {\bibfnamefont {Radouane}\
  \bibnamefont {Gannouji}}, \bibinfo {author} {\bibfnamefont {Lavrentios}\
  \bibnamefont {Kazantzidis}}, \bibinfo {author} {\bibfnamefont {Leandros}\
  \bibnamefont {Perivolaropoulos}}, \ and\ \bibinfo {author} {\bibfnamefont
  {David}\ \bibnamefont {Polarski}},\ }\bibfield  {title} {\enquote {\bibinfo
  {title} {{Consistency of modified gravity with a decreasing $G_{\rm eff}(z)$
  in a $\Lambda$CDM background}},}\ }\href {\doibase
  10.1103/PhysRevD.98.104044} {\bibfield  {journal} {\bibinfo  {journal} {Phys.
  Rev. D}\ }\textbf {\bibinfo {volume} {98}},\ \bibinfo {pages} {104044}
  (\bibinfo {year} {2018})},\ \Eprint {http://arxiv.org/abs/1809.07034}
  {arXiv:1809.07034 [gr-qc]} \BibitemShut {NoStop}%
\bibitem [{\citenamefont {Gannouji}\ \emph {et~al.}(2020)\citenamefont
  {Gannouji}, \citenamefont {Perivolaropoulos}, \citenamefont {Polarski},\ and\
  \citenamefont {Skara}}]{Gannouji:2020ylf}%
  \BibitemOpen
  \bibfield  {author} {\bibinfo {author} {\bibfnamefont {Radouane}\
  \bibnamefont {Gannouji}}, \bibinfo {author} {\bibfnamefont {Leandros}\
  \bibnamefont {Perivolaropoulos}}, \bibinfo {author} {\bibfnamefont {David}\
  \bibnamefont {Polarski}}, \ and\ \bibinfo {author} {\bibfnamefont {Foteini}\
  \bibnamefont {Skara}},\ }\bibfield  {title} {\enquote {\bibinfo {title}
  {{Weak gravity on a $\Lambda$CDM background}},}\ }\href@noop {} {\  (\bibinfo
  {year} {2020})},\ \Eprint {http://arxiv.org/abs/2011.01517} {arXiv:2011.01517
  [gr-qc]} \BibitemShut {NoStop}%
\bibitem [{\citenamefont {Amendola}\ \emph {et~al.}(2018)\citenamefont
  {Amendola}, \citenamefont {Kunz}, \citenamefont {Saltas},\ and\ \citenamefont
  {Sawicki}}]{Amendola:2017orw}%
  \BibitemOpen
  \bibfield  {author} {\bibinfo {author} {\bibfnamefont {Luca}\ \bibnamefont
  {Amendola}}, \bibinfo {author} {\bibfnamefont {Martin}\ \bibnamefont {Kunz}},
  \bibinfo {author} {\bibfnamefont {Ippocratis~D.}\ \bibnamefont {Saltas}}, \
  and\ \bibinfo {author} {\bibfnamefont {Ignacy}\ \bibnamefont {Sawicki}},\
  }\bibfield  {title} {\enquote {\bibinfo {title} {{Fate of Large-Scale
  Structure in Modified Gravity After GW170817 and GRB170817A}},}\ }\href
  {\doibase 10.1103/PhysRevLett.120.131101} {\bibfield  {journal} {\bibinfo
  {journal} {Phys. Rev. Lett.}\ }\textbf {\bibinfo {volume} {120}},\ \bibinfo
  {pages} {131101} (\bibinfo {year} {2018})},\ \Eprint
  {http://arxiv.org/abs/1711.04825} {arXiv:1711.04825 [astro-ph.CO]}
  \BibitemShut {NoStop}%
\bibitem [{\citenamefont
  {Perivolaropoulos}(2005{\natexlab{a}})}]{Perivolaropoulos:2004yr}%
  \BibitemOpen
  \bibfield  {author} {\bibinfo {author} {\bibfnamefont {Leandros}\
  \bibnamefont {Perivolaropoulos}},\ }\bibfield  {title} {\enquote {\bibinfo
  {title} {{Constraints on linear negative potentials in quintessence and
  phantom models from recent supernova data}},}\ }\href {\doibase
  10.1103/PhysRevD.71.063503} {\bibfield  {journal} {\bibinfo  {journal} {Phys.
  Rev. D}\ }\textbf {\bibinfo {volume} {71}},\ \bibinfo {pages} {063503}
  (\bibinfo {year} {2005}{\natexlab{a}})},\ \Eprint
  {http://arxiv.org/abs/astro-ph/0412308} {arXiv:astro-ph/0412308} \BibitemShut
  {NoStop}%
\bibitem [{\citenamefont {Nesseris}\ and\ \citenamefont
  {Perivolaropoulos}(2007)}]{Nesseris:2006er}%
  \BibitemOpen
  \bibfield  {author} {\bibinfo {author} {\bibfnamefont {S.}~\bibnamefont
  {Nesseris}}\ and\ \bibinfo {author} {\bibfnamefont {Leandros}\ \bibnamefont
  {Perivolaropoulos}},\ }\bibfield  {title} {\enquote {\bibinfo {title}
  {{Crossing the Phantom Divide: Theoretical Implications and Observational
  Status}},}\ }\href {\doibase 10.1088/1475-7516/2007/01/018} {\bibfield
  {journal} {\bibinfo  {journal} {JCAP}\ }\textbf {\bibinfo {volume} {01}},\
  \bibinfo {pages} {018} (\bibinfo {year} {2007})},\ \Eprint
  {http://arxiv.org/abs/astro-ph/0610092} {arXiv:astro-ph/0610092} \BibitemShut
  {NoStop}%
\bibitem [{\citenamefont
  {Perivolaropoulos}(2005{\natexlab{b}})}]{Perivolaropoulos:2005yv}%
  \BibitemOpen
  \bibfield  {author} {\bibinfo {author} {\bibfnamefont {Leandros}\
  \bibnamefont {Perivolaropoulos}},\ }\bibfield  {title} {\enquote {\bibinfo
  {title} {{Crossing the phantom divide barrier with scalar tensor
  theories}},}\ }\href {\doibase 10.1088/1475-7516/2005/10/001} {\bibfield
  {journal} {\bibinfo  {journal} {JCAP}\ }\textbf {\bibinfo {volume} {10}},\
  \bibinfo {pages} {001} (\bibinfo {year} {2005}{\natexlab{b}})},\ \Eprint
  {http://arxiv.org/abs/astro-ph/0504582} {arXiv:astro-ph/0504582} \BibitemShut
  {NoStop}%
\bibitem [{\citenamefont {Caldwell}\ \emph {et~al.}(2003)\citenamefont
  {Caldwell}, \citenamefont {Kamionkowski},\ and\ \citenamefont
  {Weinberg}}]{Caldwell:2003vq}%
  \BibitemOpen
  \bibfield  {author} {\bibinfo {author} {\bibfnamefont {Robert~R.}\
  \bibnamefont {Caldwell}}, \bibinfo {author} {\bibfnamefont {Marc}\
  \bibnamefont {Kamionkowski}}, \ and\ \bibinfo {author} {\bibfnamefont
  {Nevin~N.}\ \bibnamefont {Weinberg}},\ }\bibfield  {title} {\enquote
  {\bibinfo {title} {{Phantom energy and cosmic doomsday}},}\ }\href {\doibase
  10.1103/PhysRevLett.91.071301} {\bibfield  {journal} {\bibinfo  {journal}
  {Phys. Rev. Lett.}\ }\textbf {\bibinfo {volume} {91}},\ \bibinfo {pages}
  {071301} (\bibinfo {year} {2003})},\ \Eprint
  {http://arxiv.org/abs/astro-ph/0302506} {arXiv:astro-ph/0302506} \BibitemShut
  {NoStop}%
\bibitem [{\citenamefont {Nesseris}\ and\ \citenamefont
  {Perivolaropoulos}(2004)}]{Nesseris:2004uj}%
  \BibitemOpen
  \bibfield  {author} {\bibinfo {author} {\bibfnamefont {S.}~\bibnamefont
  {Nesseris}}\ and\ \bibinfo {author} {\bibfnamefont {Leandros}\ \bibnamefont
  {Perivolaropoulos}},\ }\bibfield  {title} {\enquote {\bibinfo {title} {{The
  Fate of bound systems in phantom and quintessence cosmologies}},}\ }\href
  {\doibase 10.1103/PhysRevD.70.123529} {\bibfield  {journal} {\bibinfo
  {journal} {Phys. Rev. D}\ }\textbf {\bibinfo {volume} {70}},\ \bibinfo
  {pages} {123529} (\bibinfo {year} {2004})},\ \Eprint
  {http://arxiv.org/abs/astro-ph/0410309} {arXiv:astro-ph/0410309} \BibitemShut
  {NoStop}%
\bibitem [{\citenamefont {Elizalde}\ \emph {et~al.}(2004)\citenamefont
  {Elizalde}, \citenamefont {Nojiri},\ and\ \citenamefont
  {Odintsov}}]{Elizalde:2004mq}%
  \BibitemOpen
  \bibfield  {author} {\bibinfo {author} {\bibfnamefont {Emilio}\ \bibnamefont
  {Elizalde}}, \bibinfo {author} {\bibfnamefont {Shin'ichi}\ \bibnamefont
  {Nojiri}}, \ and\ \bibinfo {author} {\bibfnamefont {Sergei~D.}\ \bibnamefont
  {Odintsov}},\ }\bibfield  {title} {\enquote {\bibinfo {title} {{Late-time
  cosmology in (phantom) scalar-tensor theory: Dark energy and the cosmic
  speed-up}},}\ }\href {\doibase 10.1103/PhysRevD.70.043539} {\bibfield
  {journal} {\bibinfo  {journal} {Phys. Rev. D}\ }\textbf {\bibinfo {volume}
  {70}},\ \bibinfo {pages} {043539} (\bibinfo {year} {2004})},\ \Eprint
  {http://arxiv.org/abs/hep-th/0405034} {arXiv:hep-th/0405034} \BibitemShut
  {NoStop}%
\bibitem [{\citenamefont {Beutler}\ \emph {et~al.}(2011)\citenamefont
  {Beutler}, \citenamefont {Blake}, \citenamefont {Colless}, \citenamefont
  {Jones}, \citenamefont {Staveley-Smith}, \citenamefont {Campbell},
  \citenamefont {Parker}, \citenamefont {Saunders},\ and\ \citenamefont
  {Watson}}]{Beutler:2011hx}%
  \BibitemOpen
  \bibfield  {author} {\bibinfo {author} {\bibfnamefont {Florian}\ \bibnamefont
  {Beutler}}, \bibinfo {author} {\bibfnamefont {Chris}\ \bibnamefont {Blake}},
  \bibinfo {author} {\bibfnamefont {Matthew}\ \bibnamefont {Colless}}, \bibinfo
  {author} {\bibfnamefont {D.Heath}\ \bibnamefont {Jones}}, \bibinfo {author}
  {\bibfnamefont {Lister}\ \bibnamefont {Staveley-Smith}}, \bibinfo {author}
  {\bibfnamefont {Lachlan}\ \bibnamefont {Campbell}}, \bibinfo {author}
  {\bibfnamefont {Quentin}\ \bibnamefont {Parker}}, \bibinfo {author}
  {\bibfnamefont {Will}\ \bibnamefont {Saunders}}, \ and\ \bibinfo {author}
  {\bibfnamefont {Fred}\ \bibnamefont {Watson}},\ }\bibfield  {title} {\enquote
  {\bibinfo {title} {{The 6dF Galaxy Survey: Baryon Acoustic Oscillations and
  the Local Hubble Constant}},}\ }\href {\doibase
  10.1111/j.1365-2966.2011.19250.x} {\bibfield  {journal} {\bibinfo  {journal}
  {Mon. Not. Roy. Astron. Soc.}\ }\textbf {\bibinfo {volume} {416}},\ \bibinfo
  {pages} {3017--3032} (\bibinfo {year} {2011})},\ \Eprint
  {http://arxiv.org/abs/1106.3366} {arXiv:1106.3366 [astro-ph.CO]} \BibitemShut
  {NoStop}%
\bibitem [{\citenamefont {Blake}\ \emph {et~al.}(2012)\citenamefont {Blake}
  \emph {et~al.}}]{Blake:2012pj}%
  \BibitemOpen
  \bibfield  {author} {\bibinfo {author} {\bibfnamefont {Chris}\ \bibnamefont
  {Blake}} \emph {et~al.},\ }\bibfield  {title} {\enquote {\bibinfo {title}
  {{The WiggleZ Dark Energy Survey: Joint measurements of the expansion and
  growth history at z < 1}},}\ }\href {\doibase
  10.1111/j.1365-2966.2012.21473.x} {\bibfield  {journal} {\bibinfo  {journal}
  {Mon. Not. Roy. Astron. Soc.}\ }\textbf {\bibinfo {volume} {425}},\ \bibinfo
  {pages} {405--414} (\bibinfo {year} {2012})},\ \Eprint
  {http://arxiv.org/abs/1204.3674} {arXiv:1204.3674 [astro-ph.CO]} \BibitemShut
  {NoStop}%
\bibitem [{\citenamefont {de~Sainte~Agathe}\ \emph {et~al.}(2019)\citenamefont
  {de~Sainte~Agathe} \emph {et~al.}}]{Agathe:2019vsu}%
  \BibitemOpen
  \bibfield  {author} {\bibinfo {author} {\bibfnamefont {Victoria}\
  \bibnamefont {de~Sainte~Agathe}} \emph {et~al.},\ }\bibfield  {title}
  {\enquote {\bibinfo {title} {{Baryon acoustic oscillations at z = 2.34 from
  the correlations of Ly$\alpha$ absorption in eBOSS DR14}},}\ }\href {\doibase
  10.1051/0004-6361/201935638} {\bibfield  {journal} {\bibinfo  {journal}
  {Astron. Astrophys.}\ }\textbf {\bibinfo {volume} {629}},\ \bibinfo {pages}
  {A85} (\bibinfo {year} {2019})},\ \Eprint {http://arxiv.org/abs/1904.03400}
  {arXiv:1904.03400 [astro-ph.CO]} \BibitemShut {NoStop}%
\bibitem [{\citenamefont {Ross}\ \emph {et~al.}(2015)\citenamefont {Ross},
  \citenamefont {Samushia}, \citenamefont {Howlett}, \citenamefont {Percival},
  \citenamefont {Burden},\ and\ \citenamefont {Manera}}]{Ross:2014qpa}%
  \BibitemOpen
  \bibfield  {author} {\bibinfo {author} {\bibfnamefont {Ashley~J.}\
  \bibnamefont {Ross}}, \bibinfo {author} {\bibfnamefont {Lado}\ \bibnamefont
  {Samushia}}, \bibinfo {author} {\bibfnamefont {Cullan}\ \bibnamefont
  {Howlett}}, \bibinfo {author} {\bibfnamefont {Will~J.}\ \bibnamefont
  {Percival}}, \bibinfo {author} {\bibfnamefont {Angela}\ \bibnamefont
  {Burden}}, \ and\ \bibinfo {author} {\bibfnamefont {Marc}\ \bibnamefont
  {Manera}},\ }\bibfield  {title} {\enquote {\bibinfo {title} {{The clustering
  of the SDSS DR7 main Galaxy sample -- I. A 4 per cent distance measure at $z
  = 0.15$}},}\ }\href {\doibase 10.1093/mnras/stv154} {\bibfield  {journal}
  {\bibinfo  {journal} {Mon. Not. Roy. Astron. Soc.}\ }\textbf {\bibinfo
  {volume} {449}},\ \bibinfo {pages} {835--847} (\bibinfo {year} {2015})},\
  \Eprint {http://arxiv.org/abs/1409.3242} {arXiv:1409.3242 [astro-ph.CO]}
  \BibitemShut {NoStop}%
\bibitem [{\citenamefont {Anderson}\ \emph {et~al.}(2014)\citenamefont
  {Anderson} \emph {et~al.}}]{Anderson:2013zyy}%
  \BibitemOpen
  \bibfield  {author} {\bibinfo {author} {\bibfnamefont {Lauren}\ \bibnamefont
  {Anderson}} \emph {et~al.} (\bibinfo {collaboration} {BOSS}),\ }\bibfield
  {title} {\enquote {\bibinfo {title} {{The clustering of galaxies in the
  SDSS-III Baryon Oscillation Spectroscopic Survey: baryon acoustic
  oscillations in the Data Releases 10 and 11 Galaxy samples}},}\ }\href
  {\doibase 10.1093/mnras/stu523} {\bibfield  {journal} {\bibinfo  {journal}
  {Mon. Not. Roy. Astron. Soc.}\ }\textbf {\bibinfo {volume} {441}},\ \bibinfo
  {pages} {24--62} (\bibinfo {year} {2014})},\ \Eprint
  {http://arxiv.org/abs/1312.4877} {arXiv:1312.4877 [astro-ph.CO]} \BibitemShut
  {NoStop}%
\bibitem [{\citenamefont {Jimenez}\ \emph {et~al.}(2003)\citenamefont
  {Jimenez}, \citenamefont {Verde}, \citenamefont {Treu},\ and\ \citenamefont
  {Stern}}]{Jimenez2003Feb}%
  \BibitemOpen
  \bibfield  {author} {\bibinfo {author} {\bibfnamefont {Raul}\ \bibnamefont
  {Jimenez}}, \bibinfo {author} {\bibfnamefont {Licia}\ \bibnamefont {Verde}},
  \bibinfo {author} {\bibfnamefont {Tommaso}\ \bibnamefont {Treu}}, \ and\
  \bibinfo {author} {\bibfnamefont {Daniel}\ \bibnamefont {Stern}},\ }\bibfield
   {title} {\enquote {\bibinfo {title} {{Constraints on the equation of state
  of dark energy and the Hubble constant from stellar ages and the CMB}},}\
  }\href {\doibase 10.1086/376595} {\bibfield  {journal} {\bibinfo  {journal}
  {ArXiv e-prints}\ } (\bibinfo {year} {2003}),\ 10.1086/376595},\ \Eprint
  {http://arxiv.org/abs/astro-ph/0302560} {astro-ph/0302560} \BibitemShut
  {NoStop}%
\bibitem [{\citenamefont {Simon}\ \emph {et~al.}(2005)\citenamefont {Simon},
  \citenamefont {Verde},\ and\ \citenamefont {Jimenez}}]{PhysRevD.71.123001}%
  \BibitemOpen
  \bibfield  {author} {\bibinfo {author} {\bibfnamefont {Joan}\ \bibnamefont
  {Simon}}, \bibinfo {author} {\bibfnamefont {Licia}\ \bibnamefont {Verde}}, \
  and\ \bibinfo {author} {\bibfnamefont {Raul}\ \bibnamefont {Jimenez}},\
  }\bibfield  {title} {\enquote {\bibinfo {title} {Constraints on the redshift
  dependence of the dark energy potential},}\ }\href {\doibase
  10.1103/PhysRevD.71.123001} {\bibfield  {journal} {\bibinfo  {journal} {Phys.
  Rev. D}\ }\textbf {\bibinfo {volume} {71}},\ \bibinfo {pages} {123001}
  (\bibinfo {year} {2005})}\BibitemShut {NoStop}%
\bibitem [{\citenamefont {Moresco}\ \emph {et~al.}(2012)\citenamefont
  {Moresco}, \citenamefont {Cimatti}, \citenamefont {Jimenez}, \citenamefont
  {Pozzetti}, \citenamefont {Zamorani}, \citenamefont {Bolzonella},
  \citenamefont {Dunlop}, \citenamefont {Lamareille}, \citenamefont {Mignoli},
  \citenamefont {Pearce},\ and\ \citenamefont {et~al.}}]{Moresco_2012}%
  \BibitemOpen
  \bibfield  {author} {\bibinfo {author} {\bibfnamefont {M}~\bibnamefont
  {Moresco}}, \bibinfo {author} {\bibfnamefont {A}~\bibnamefont {Cimatti}},
  \bibinfo {author} {\bibfnamefont {R}~\bibnamefont {Jimenez}}, \bibinfo
  {author} {\bibfnamefont {L}~\bibnamefont {Pozzetti}}, \bibinfo {author}
  {\bibfnamefont {G}~\bibnamefont {Zamorani}}, \bibinfo {author} {\bibfnamefont
  {M}~\bibnamefont {Bolzonella}}, \bibinfo {author} {\bibfnamefont
  {J}~\bibnamefont {Dunlop}}, \bibinfo {author} {\bibfnamefont {F}~\bibnamefont
  {Lamareille}}, \bibinfo {author} {\bibfnamefont {M}~\bibnamefont {Mignoli}},
  \bibinfo {author} {\bibfnamefont {H}~\bibnamefont {Pearce}}, \ and\ \bibinfo
  {author} {\bibnamefont {et~al.}},\ }\bibfield  {title} {\enquote {\bibinfo
  {title} {Improved constraints on the expansion rate of the universe up to $z
  \sim 1.1$ from the spectroscopic evolution of cosmic chronometers},}\ }\href
  {\doibase 10.1088/1475-7516/2012/08/006} {\bibfield  {journal} {\bibinfo
  {journal} {Journal of Cosmology and Astroparticle Physics}\ }\textbf
  {\bibinfo {volume} {2012}},\ \bibinfo {pages} {006–006} (\bibinfo {year}
  {2012})}\BibitemShut {NoStop}%
\bibitem [{\citenamefont {Moresco}\ \emph {et~al.}(2016)\citenamefont
  {Moresco}, \citenamefont {Pozzetti}, \citenamefont {Cimatti}, \citenamefont
  {Jimenez}, \citenamefont {Maraston}, \citenamefont {Verde}, \citenamefont
  {Thomas}, \citenamefont {Citro}, \citenamefont {Tojeiro},\ and\ \citenamefont
  {Wilkinson}}]{Moresco2016May}%
  \BibitemOpen
  \bibfield  {author} {\bibinfo {author} {\bibfnamefont {Michele}\ \bibnamefont
  {Moresco}}, \bibinfo {author} {\bibfnamefont {Lucia}\ \bibnamefont
  {Pozzetti}}, \bibinfo {author} {\bibfnamefont {Andrea}\ \bibnamefont
  {Cimatti}}, \bibinfo {author} {\bibfnamefont {Raul}\ \bibnamefont {Jimenez}},
  \bibinfo {author} {\bibfnamefont {Claudia}\ \bibnamefont {Maraston}},
  \bibinfo {author} {\bibfnamefont {Licia}\ \bibnamefont {Verde}}, \bibinfo
  {author} {\bibfnamefont {Daniel}\ \bibnamefont {Thomas}}, \bibinfo {author}
  {\bibfnamefont {Annalisa}\ \bibnamefont {Citro}}, \bibinfo {author}
  {\bibfnamefont {Rita}\ \bibnamefont {Tojeiro}}, \ and\ \bibinfo {author}
  {\bibfnamefont {David}\ \bibnamefont {Wilkinson}},\ }\bibfield  {title}
  {\enquote {\bibinfo {title} {{A 6{\%} measurement of the Hubble parameter at
  z{$\sim$}0.45: direct evidence of the epoch of cosmic re-acceleration}},}\
  }\href {\doibase 10.1088/1475-7516/2016/05/014} {\bibfield  {journal}
  {\bibinfo  {journal} {Journal of Cosmology and Astroparticle Physics}\
  }\textbf {\bibinfo {volume} {2016}},\ \bibinfo {pages} {014} (\bibinfo {year}
  {2016})}\BibitemShut {NoStop}%
\bibitem [{\citenamefont {Stern}\ \emph {et~al.}(2010)\citenamefont {Stern},
  \citenamefont {Jimenez}, \citenamefont {Verde}, \citenamefont
  {Kamionkowski},\ and\ \citenamefont {Stanford}}]{Stern2010Feb}%
  \BibitemOpen
  \bibfield  {author} {\bibinfo {author} {\bibfnamefont {Daniel}\ \bibnamefont
  {Stern}}, \bibinfo {author} {\bibfnamefont {Raul}\ \bibnamefont {Jimenez}},
  \bibinfo {author} {\bibfnamefont {Licia}\ \bibnamefont {Verde}}, \bibinfo
  {author} {\bibfnamefont {Marc}\ \bibnamefont {Kamionkowski}}, \ and\ \bibinfo
  {author} {\bibfnamefont {S.~Adam}\ \bibnamefont {Stanford}},\ }\bibfield
  {title} {\enquote {\bibinfo {title} {{Cosmic chronometers: constraining the
  equation of state of dark energy. I: H(z) measurements}},}\ }\href {\doibase
  10.1088/1475-7516/2010/02/008} {\bibfield  {journal} {\bibinfo  {journal}
  {Journal of Cosmology and Astroparticle Physics}\ }\textbf {\bibinfo {volume}
  {2010}},\ \bibinfo {pages} {008} (\bibinfo {year} {2010})}\BibitemShut
  {NoStop}%
\bibitem [{\citenamefont {Zhang}\ \emph {et~al.}(2012)\citenamefont {Zhang},
  \citenamefont {Zhang}, \citenamefont {Yuan}, \citenamefont {Liu},
  \citenamefont {Zhang},\ and\ \citenamefont {Sun}}]{Zhang2012Jul}%
  \BibitemOpen
  \bibfield  {author} {\bibinfo {author} {\bibfnamefont {Cong}\ \bibnamefont
  {Zhang}}, \bibinfo {author} {\bibfnamefont {Han}\ \bibnamefont {Zhang}},
  \bibinfo {author} {\bibfnamefont {Shuo}\ \bibnamefont {Yuan}}, \bibinfo
  {author} {\bibfnamefont {Siqi}\ \bibnamefont {Liu}}, \bibinfo {author}
  {\bibfnamefont {Tong-Jie}\ \bibnamefont {Zhang}}, \ and\ \bibinfo {author}
  {\bibfnamefont {Yan-Chun}\ \bibnamefont {Sun}},\ }\bibfield  {title}
  {\enquote {\bibinfo {title} {{Four New Observational $H(z)$ Data From
  Luminous Red Galaxies of Sloan Digital Sky Survey Data Release Seven}},}\
  }\href {\doibase 10.1088/1674--4527/14/10/002} {\bibfield  {journal}
  {\bibinfo  {journal} {ArXiv e-prints}\ } (\bibinfo {year} {2012}),\
  10.1088/1674--4527/14/10/002},\ \Eprint {http://arxiv.org/abs/1207.4541}
  {1207.4541} \BibitemShut {NoStop}%
\bibitem [{\citenamefont {Moresco}(2015)}]{Moresco2015Jun}%
  \BibitemOpen
  \bibfield  {author} {\bibinfo {author} {\bibfnamefont {Michele}\ \bibnamefont
  {Moresco}},\ }\bibfield  {title} {\enquote {\bibinfo {title} {{Raising the
  bar: new constraints on the Hubble parameter with cosmic chronometers at z
  {$\sim$} 2}},}\ }\href {\doibase 10.1093/mnrasl/slv037} {\bibfield  {journal}
  {\bibinfo  {journal} {Monthly Notices of the Royal Astronomical Society:
  Letters}\ }\textbf {\bibinfo {volume} {450}},\ \bibinfo {pages} {L16--L20}
  (\bibinfo {year} {2015})}\BibitemShut {NoStop}%
\bibitem [{\citenamefont {Chuang}\ and\ \citenamefont
  {Wang}(2013)}]{Chuang2013Oct}%
  \BibitemOpen
  \bibfield  {author} {\bibinfo {author} {\bibfnamefont {Chia-Hsun}\
  \bibnamefont {Chuang}}\ and\ \bibinfo {author} {\bibfnamefont {Yun}\
  \bibnamefont {Wang}},\ }\bibfield  {title} {\enquote {\bibinfo {title}
  {{Modelling the anisotropic two-point galaxy correlation function on small
  scales and single-probe measurements of H(z), DA(z) and f(z){$\sigma$}8(z)
  from the Sloan Digital Sky Survey DR7 luminous red galaxies}},}\ }\href
  {\doibase 10.1093/mnras/stt1290} {\bibfield  {journal} {\bibinfo  {journal}
  {Monthly Notices of the Royal Astronomical Society}\ }\textbf {\bibinfo
  {volume} {435}},\ \bibinfo {pages} {255--262} (\bibinfo {year}
  {2013})}\BibitemShut {NoStop}%
\bibitem [{\citenamefont {Blake~et al.}(2012)}]{Blake2012Sep}%
  \BibitemOpen
  \bibfield  {author} {\bibinfo {author} {\bibfnamefont {Chris}\ \bibnamefont
  {Blake~et al.}},\ }\bibfield  {title} {\enquote {\bibinfo {title} {{The
  WiggleZ Dark Energy Survey: joint measurements of the expansion and growth
  history at z {$<$} 1}},}\ }\href {\doibase 10.1111/j.1365-2966.2012.21473.x}
  {\bibfield  {journal} {\bibinfo  {journal} {Monthly Notices of the Royal
  Astronomical Society}\ }\textbf {\bibinfo {volume} {425}},\ \bibinfo {pages}
  {405--414} (\bibinfo {year} {2012})}\BibitemShut {NoStop}%
\bibitem [{\citenamefont {Delubac~et al.}(2015)}]{Delubac2015Feb}%
  \BibitemOpen
  \bibfield  {author} {\bibinfo {author} {\bibfnamefont
  {Timoth{\ifmmode\acute{e}\else\'{e}\fi}e}\ \bibnamefont {Delubac~et al.}},\
  }\bibfield  {title} {\enquote {\bibinfo {title} {{Baryon acoustic
  oscillations in the Ly{$\alpha$} forest of BOSS DR11 quasars}},}\ }\href
  {\doibase 10.1051/0004-6361/201423969} {\bibfield  {journal} {\bibinfo
  {journal} {Astronomy {\&} Astrophysics}\ }\textbf {\bibinfo {volume} {574}},\
  \bibinfo {pages} {A59} (\bibinfo {year} {2015})}\BibitemShut {NoStop}%
\bibitem [{\citenamefont {Gaztañaga}\ \emph {et~al.}(2009)\citenamefont
  {Gaztañaga}, \citenamefont {Cabré},\ and\ \citenamefont
  {Hui}}]{Gazta_aga_2009}%
  \BibitemOpen
  \bibfield  {author} {\bibinfo {author} {\bibfnamefont {Enrique}\ \bibnamefont
  {Gaztañaga}}, \bibinfo {author} {\bibfnamefont {Anna}\ \bibnamefont
  {Cabré}}, \ and\ \bibinfo {author} {\bibfnamefont {Lam}\ \bibnamefont
  {Hui}},\ }\bibfield  {title} {\enquote {\bibinfo {title} {Clustering of
  luminous red galaxies - iv. baryon acoustic peak in the line-of-sight
  direction and a direct measurement of h(z)},}\ }\href {\doibase
  10.1111/j.1365-2966.2009.15405.x} {\bibfield  {journal} {\bibinfo  {journal}
  {Monthly Notices of the Royal Astronomical Society}\ }\textbf {\bibinfo
  {volume} {399}},\ \bibinfo {pages} {1663–1680} (\bibinfo {year}
  {2009})}\BibitemShut {NoStop}%
\bibitem [{\citenamefont {Samushia~et al.}(2013)}]{Samushia2013Feb}%
  \BibitemOpen
  \bibfield  {author} {\bibinfo {author} {\bibfnamefont {Lado}\ \bibnamefont
  {Samushia~et al.}},\ }\bibfield  {title} {\enquote {\bibinfo {title} {{The
  clustering of galaxies in the SDSS-III DR9 Baryon Oscillation Spectroscopic
  Survey: testing deviations from {$\Lambda$} and general relativity using
  anisotropic clustering of galaxies}},}\ }\href {\doibase
  10.1093/mnras/sts443} {\bibfield  {journal} {\bibinfo  {journal} {Monthly
  Notices of the Royal Astronomical Society}\ }\textbf {\bibinfo {volume}
  {429}},\ \bibinfo {pages} {1514--1528} (\bibinfo {year} {2013})}\BibitemShut
  {NoStop}%
\bibitem [{\citenamefont {Busca~et al.}(2013)}]{Busca2013Apr}%
  \BibitemOpen
  \bibfield  {author} {\bibinfo {author} {\bibfnamefont {N.~G.}\ \bibnamefont
  {Busca~et al.}},\ }\bibfield  {title} {\enquote {\bibinfo {title} {{Baryon
  acoustic oscillations in the Ly{$\alpha$} forest of BOSS quasars}},}\ }\href
  {\doibase 10.1051/0004-6361/201220724} {\bibfield  {journal} {\bibinfo
  {journal} {Astronomy {\&} Astrophysics}\ }\textbf {\bibinfo {volume} {552}},\
  \bibinfo {pages} {A96} (\bibinfo {year} {2013})}\BibitemShut {NoStop}%
\bibitem [{\citenamefont {et~al.}(2014)}]{Font_Ribera_2014}%
  \BibitemOpen
  \bibfield  {author} {\bibinfo {author} {\bibfnamefont {Andreu Font-Ribera}\
  \bibnamefont {et~al.}},\ }\bibfield  {title} {\enquote {\bibinfo {title}
  {Quasar-lyman $\alpha$ forest cross-correlation from boss dr11: Baryon
  acoustic oscillations},}\ }\href {\doibase 10.1088/1475-7516/2014/05/027}
  {\bibfield  {journal} {\bibinfo  {journal} {Journal of Cosmology and
  Astroparticle Physics}\ }\textbf {\bibinfo {volume} {2014}},\ \bibinfo
  {pages} {027–027} (\bibinfo {year} {2014})}\BibitemShut {NoStop}%
\end{thebibliography}%

\end{document}